\documentclass{aa}  

\usepackage{graphicx}
\usepackage{txfonts}
\usepackage{xcolor}
\usepackage[colorlinks=true,citecolor=blue,linkcolor=blue]{hyperref}

\newcommand{\xmm}{\textit{XMM-Newton}\xspace}

\newcommand{\integral}{\textit{INTEGRAL}\xspace}
\newcommand{\nustar}{\textit{NuSTAR}\xspace}
\newcommand{\nicer}{NICER\xspace}
\newcommand{\ixpe}{IXPE\xspace}

\newcommand{\fluxcgs}{erg~s$^{-1}$~cm$^{-2}$\xspace}
\newcommand{\lumcgs}{erg~s$^{-1}$\xspace}
\newcommand{\source}{GX~3+1\xspace}

\def\diskbb{\texttt{diskbb}\xspace}
\def\thcomp{\texttt{thcomp}\xspace}
\def\bbodyrad{\texttt{bbodyrad}\xspace}

\def\polconst{\texttt{polconst}\xspace}

\def\relxillns{\texttt{relxillNS}\xspace}

\xspaceaddexceptions{+ - *}

\graphicspath{{./}{Figures/}}

\begin{document} 

\title{First spectropolarimetric observation of the neutron star low-mass X-ray binary GX~3+1}

\author{Andrea Gnarini \inst{\ref{in:UniRoma3}}\fnmsep\thanks{E-mail: andrea.gnarini@uniroma3.it}
\and Ruben Farinelli\inst{\ref{in:INAF-OAS}}
\and Francesco Ursini \inst{\ref{in:UniRoma3}}
\and Stefano Bianchi \inst{\ref{in:UniRoma3}}
\and Fiamma Capitanio \inst{\ref{in:INAF-IAPS}}
\and Giorgio Matt \inst{\ref{in:UniRoma3}}
\and Mason Ng\inst{\ref{in:McGill},\ref{in:McGill-TSI}}
\and Antonella Tarana \inst{\ref{in:INAF-IAPS}}
\and Anna Bobrikova\inst{\ref{in:UTU}}
\and Massimo Cocchi\inst{\ref{in:INAF-OAC}}
\and Sergio Fabiani \inst{\ref{in:INAF-IAPS}}
\and Philip Kaaret\inst{\ref{in:NASA.MSFC}}
\and Juri Poutanen\inst{\ref{in:UTU}}
\and Swati Ravi\inst{\ref{in:MIT}}
}

\institute{Dipartimento di Matematica e Fisica, Università degli Studi Roma Tre, via della Vasca Navale 84, I-00146 Roma, Italy \label{in:UniRoma3}
\and
INAF -- Osservatorio di Astrofisica e Scienza dello Spazio, Via P. Gobetti 101, 40129 Bologna, Italy \label{in:INAF-OAS}
\and
INAF -- Istituto di Astrofisica e Planetologia Spaziali, Via del Fosso del Cavaliere 100, 00133 Roma, Italy \label{in:INAF-IAPS}
\and 
Department of Physics, McGill University, 3600 rue University, Montr\'{e}al, QC H3A 2T8, Canada \label{in:McGill}
\and
Trottier Space Institute, McGill University, 3550 rue University, Montr\'{e}al, QC H3A 2A7, Canada \label{in:McGill-TSI}
\and 
Department of Physics and Astronomy, FI-20014 University of Turku, Finland \label{in:UTU}
\and
INAF Osservatorio Astronomico di Cagliari, via della Scienza 5, I-09047 Selargius (CA), Italy \label{in:INAF-OAC}
\and
NASA Marshall Space Flight Center, Huntsville, AL 35812, USA \label{in:NASA.MSFC}
\and
MIT Kavli Institute for Astrophysics and Space Research, Massachusetts Institute of Technology, Cambridge, MA 02139, USA \label{in:MIT}
}

\date{Received XXX; accepted YYY}

% \abstract{}{}{}{}{} 
% 5 {} token are mandatory

\abstract{ 
    We report the first simultaneous X-ray spectropolarimetric observation of the bright atoll neutron star low-mass X-ray binary \source, performed by the Imaging X-ray Polarimetry Explorer (\ixpe) joint with \nicer and \nustar. The source does not exhibit significant polarization in the 2--8 keV energy band, with an upper limit of 1.3\% at a 99\% confidence level on the polarization degree. The observed spectra can be well described by a combination of thermal disk emission, the hard Comptonization component, and reflected photons off the accretion disk. In particular, from the broad Fe K$\alpha$ line profile, we were able to determine the inclination of the system ($i \approx 36$\degr), which is crucial for comparing the observed polarization with theoretical models. Both the spectral and polarization properties of \source are consistent with those of other atoll sources observed by \ixpe. Therefore, we may expect a similar geometrical configuration for the accreting system and the hot Comptonizing region. The low polarization is also consistent with the low inclination of the system.
}


\maketitle

%%%%%%%%%%%%%%%%% INTRODUCTION %%%%%%%%%%%%%%%%%%

\section{Introduction}

Weakly-magnetized neutron stars in low-mass X-ray binaries (NS-LMXBs) are among the brightest sources in the X-ray sky and are perfect laboratories for studying the properties of accretion in the strong gravity regime with X-ray polarimetry. Matter accretes onto the NS surface through Roche-lobe overflow, typically from a late main-sequence or evolved degenerate companion star. The surface of the NS stops the accretion flow, creating a boundary (BL) or spreading layer (SL) between the inner edge of the accretion disk and the NS, which extends to higher latitudes as the accretion rate increases \citep{Inogamov.Sunyaev.1999,Popham.Sunyaev.2001,Suleimanov.Poutanen.2006}.

According to their joint timing and spectral properties in the X-ray band, NS-LMXBs are traditionally divided into a few broad categories according to their tracks in the Hard-color/Soft-color diagrams (CCDs) or hardness-intensity diagrams (HIDs; \citealt{Hasinger.VanDerKlis.1989,VanDerKlis.1989,VanDerKlis.2006}:  Atolls ($L \sim 10^{36}$\,\lumcgs) show less extended and more compact tracks in a rounded single spot in the hard region of the CCD, termed the island state, or in a banana-shaped branch for bright Atolls ($L = 10^{37}-10^{38}$\,\lumcgs; \citealt{VanDerKlis.1995}); $Z$-sources ($L > 10^{38}$\,\lumcgs) exhibit a wide $Z$-like three-branch pattern in the CCD \citep{VanDerKlis.2006,Kuulkers.etAl.1994,Migliari.Fender.2006}. At lower luminosities, Atoll sources generally trace isolated patches on timescales from days to weeks, whereas the motion is faster at higher luminosities when they move along the banana branch, sometimes showing secular motion that does not influence the variability of the source \citep{Migliari.Fender.2006}.

X-ray spectra of NS-LMXBs are typically modeled with a thermal component dominating at lower energies (below 1--2 keV) plus a harder Comptonized emission. Thermal emission may be related to the direct emission of the accretion disk \citep{Mitsuda.etAl.1984,Mitsuda.etAl.1989} or of the NS surface, while hard emission is due to inverse Compton scattering of the soft photons by a hot electron plasma, whose geometric configuration and physical properties are still not clear. In addition, the accretion disk can reprocess the primary X-ray continuum, leading to a reflection component that usually includes fluorescent emission lines (\citealt{DiSalvo.etAl.2009,Miller.etAl.2013,Mondal.etAl.2017,Ludlam.etAl.2019,Iaria.etAl.2020}; see also review by \citealt{Ludlam.2024}). X-ray polarimetry, along with timing and spectral observations, could be crucial for determining the geometry and physical characteristics of the system.

The Imaging X-ray Polarimetry Explorer (\ixpe; \citealt{Weisskopf.etAl.2016,Weisskopf.2022}), successfully launched on 2021 December 9, has introduced a powerful new instrument for studying the physical properties of various types of X-ray astronomical objects. \ixpe features three X-ray telescopes that use polarization-sensitive imaging detectors (the gas-pixel detectors; \citealt{Costa.2001}) to measure the Stokes parameters of the X-ray radiation, operating in the 2--8 keV energy range. 

Being bright sources, \ixpe has observed several NS-LMXBs so far (see also the review by \citealt{Ursini.2024.Review}): out of the twelve NS-LMXBs observed so far (seven $Z$-sources, five Atolls, and two peculiar sources), Atolls are generally less polarized than $Z$-sources in the 2--8 keV range \citep{Capitanio.etAl.2023,Ursini.etAl.2023,DiMarco.etAl.2023.4U,Saade.etAl.2024,Ursini.etAl.2024}, but exhibit an increasing trend of the polarization degree with energy. The polarization signal in most observed NS-LMXBs is related to the hard component, characterized by a higher polarization compared to the soft thermal emission \citep{Farinelli.etAl.2023,Cocchi.etAl.2023,Ursini.etAl.2023,LaMonaca.etAl.2024}. In the majority of NS-LMXBs observed by \ixpe, the polarization of the hard component seems to originate from the BL or SL, plus the contribution of reflection: these soft photons reflected off the accretion disk may significantly contribute to the polarization signal \citep{Lapidus.Sunyaev.1985,Matt.1993}, despite their relatively minor contribution to the overall flux. A peculiar behavior was observed for the ultracompact Atoll 4U~1820--303, which exhibited a rapid increase in polarization up to $10\% \pm 2\%$ in the 7--8 keV band \citep{DiMarco.etAl.2023.4U}, much higher than expected for typical BL or SL configurations \citep{Gnarini.etAl.2022,Capitanio.etAl.2023,Ursini.etAl.2023,Farinelli.etAl.2024,Bobrikova.etAl.2024.SL}. Therefore, a different shape of the Comptonizing region or an additional contribution, for example from outflows, is required to explain this high polarization.

\source is a persistent and bright Atoll discovered by \cite{Bowyer.etAl.1965} and identified as an accreting NS in a LMXB system from the detection of Type I X-ray bursts with typical timescales of tens of seconds \citep{Makishima.etAl.1983,Kuulkers.VanDerKlis.2000,denHartog.etAl.2003,Chenevez.etAl.2006} and exhibits superbursts of longer duration (up to hours) due to carbon burning \citep{Kuulkers.2002}. Assuming that type I X-ray bursts are Eddington limited, the estimated upper limit to the source distance is $\sim 6.5$ kpc \citep{Galloway.etAl.2008}, consistent with the results obtained from a double peak photospheric radius expansion burst \citep{Kuulkers.VanDerKlis.2000}. The near-infrared (NIR) counterpart to \source was discovered by \cite{VanDenBerg.etAl.2014} as a star with $15.8 \pm 0.1$ mag, suggesting that the donor star may not be a late-type giant with the NIR dominated by the emission of the outer accretion disk. \source is among the brightest persistent Atoll sources associated with the bulge component of the Milky Way. These sources are typically characterized by high X-ray luminosity ($10^{37}-10^{38}$\,\lumcgs), with moderate and irregular flux variations. Similarly to the bright atolls GX~9+1 and GX~9+9, \source exhibits strong long-term X-ray flux modulations, with a timescale of $\sim 6$ yr \citep{Kotze.Charles.2010,Durant.etAl.2010}, likely due to variations in the mass-transfer rate. Moreover, this NS-LMXB was observed only in its banana branch \citep{Lewin.etAl.1987,Hasinger.VanDerKlis.1989,Asai.etAl.1993,Pintore.etAl.2015,Thomas.etAl.2023}, without any evidence of island state in its CCD or HID. 

The X-ray spectrum of \source can be well described by the combination of a soft thermal component, most likely related to the accretion disk, plus a Comptonized component due to the hot electron population in the corona \citep{Oosterbroek.etAl.2001,Mainardi.etAl.2010,Seifina.Titarchuk.2012,Piraino.etAl.2012,Pintore.etAl.2015,Ludlam.etAl.2019}. Several K$\alpha$ broad emission lines of Ar, Ca, and Fe were also detected in \source spectra \citep{Oosterbroek.etAl.2001,Piraino.etAl.2012}. These features arise from the reprocessing of X-ray photons illuminating the accretion disk. In particular, the location of the inner disk radius and the inclination of the system can be estimated from the relativistically broadened Fe K$\alpha$ line profile: from \xmm and \integral observations, the inclination is $\sim 35\degr$ with the inner edge of the accretion disk located at $\sim 10$ $R_{\rm g}$ (where $R_{\rm g} = GM/c^2$; \citealt{Pintore.etAl.2015}), while it appears to be closer to the innermost stable circular orbit (ISCO) in a \nustar observation, with an inclination of $27\degr - 31\degr$ \citep{Ludlam.etAl.2019}.

The paper is structured as follows. In Sect. \ref{sec:Observations}, we describe the \ixpe, \nustar, and \nicer observations and data reduction. In Sect. \ref{sec:Data.Analysis}, we present the X-ray spectropolarimetric analysis. Finally, in Sect. \ref{sec:Discussion} and \ref{sec:Conclusions}, we discuss the results and the main conclusions.  

%%%%%%%%%%%%%%%%% OBSERVATIONS %%%%%%%%%%%%%%%%%%

\section{Observations and data reduction}\label{sec:Observations}

\begin{table}
\caption{Log of the observations of \source.}             
\label{table:Obs}      
\centering                                     
\begin{tabular}{llcc}         
\hline\hline
\noalign{\smallskip}
  Satellite & Obs. ID & Start Date (UTC) & Exp (ks) \\   
\hline     
\noalign{\smallskip}
  \ixpe & 03004101 & 2024-08-16 07:26:28 & 47.7 \\
  \nustar & 31001012002 & 2024-08-16 16:56:09 & 34.3 \\
  \nicer & 7700040101 & 2024-08-16 15:35:51 & 1.0 \\
  \nicer & 7700040102 & 2024-08-17 00:53:20 & 0.2 \\
\hline
\end{tabular}
\end{table}

\subsection{\ixpe}

\ixpe observed \source from 2024 August 16 07:26:28 UT to August 17 08:24:32 UT for a total exposure time of 47.7 ks (see Table \ref{table:Obs}). We extracted time-resolved spectral and polarimetric data files for each detector unit (DU) with the standard \textsc{ftools} procedure\footnote{\url{https://heasarc.gsfc.nasa.gov/docs/ixpe/analysis/IXPE-SOC-DOC-009D_UG-Software.pdf}} \citep[HEASoft v6.33;][]{Heasarc} and the latest available calibration files (CALDB v.20240125). The data analysis was performed considering the weighted analysis method presented in \cite{DiMarco.etAl.2022} with the parameter \texttt{stokes=Neff} in \textsc{xselect} (see also \citealt{Baldini.etAl.2022}). We extracted the source spectra and light curves from a circular region of 120\arcsec\ radius centered on the source, with radius derived through an iterative procedure to maximize the signal-to-noise ratio (S/N) in the 2--8 keV band, similar to the approach described in \cite{Piconcelli.etAl.2004}. The background was extracted from an annular region centered on the source with an inner and outer radius of 180\arcsec\ and 240\arcsec respectively, but it was not subtracted during the analysis, following the prescription by \cite{DiMarco.etAl.2023} for bright sources. However, we verified that the subtraction of the background has a minimal impact on the results, particularly on the polarimetric measurements. The ancillary response file (ARF) and modulation response file (MRF) for each DU was generated using the \texttt{ixpecalcarf} task, considering the same extraction radius used for the source region.
\begin{figure}
    \centering
    \includegraphics[width=0.475\textwidth]{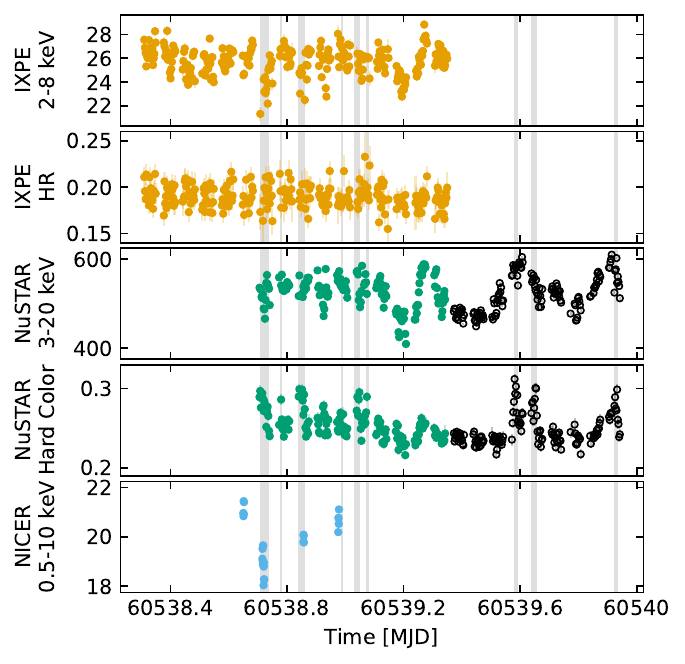}
    \caption{\ixpe, \nustar, and \nicer light curves of \source (count\,s$^{-1}$). The second panel shows the \ixpe hardness ratio (5--8 keV/3--5 keV). The fourth panel shows the \nustar hard color (10--20 keV/6--10 keV). Empty black circles denote the \nustar points that are not simultaneous with the \ixpe observation. The gray regions correspond to the time intervals during which the source moves in the UB branch. For \ixpe and \nustar, we considered time bins of 200 s, whereas for \nicer we opted for 50 s bins.}
    \label{fig:LC}
\end{figure}

\subsection{\nustar}

The Nuclear Spectroscopic Telescope Array (\nustar; \citealt{Harrison.etAl.2013}) observed \source with its X-ray telescopes on focal plane module A (FPMA) and B (FPMB) from 2024 August 16 16:56:09 UT to August 17 22:56:09 UT for a net exposure time of 34.3 ks (see Table \ref{table:Obs}). We extracted clean event files using the latest calibration files (CALDB v.20240812) and the standard \texttt{nupipeline} task of the \nustar Data Analysis Software (\textsc{nustardas} v.2.1.4) with the keyword \texttt{statusexpr="(STATUS==b0000xxx00xxxx000)\&\&(SHIELD= =0)"} due to the source brightness ($> 100$ counts s$^{-1}$). The source spectrum and light curves for each FPM were extracted using the \texttt{nuproducts} task from a circular region centered on the source with 120\arcsec\ radius. The extraction radii of the source regions were calculated with the same procedure of maximization of the S/N used for \ixpe data reduction (see also \citealt{Piconcelli.etAl.2004,Marinucci.etAl.2022,Ursini.etAl.2023.CirGal}). Since the background is not negligible at all energies, we performed background subtraction. For both FPMs, the background was extracted from a circular region with 60\arcsec\ radius sufficiently far from the source. For the spectropolarimetric analysis, we considered the \nustar data in the 3--25 keV range, since the background starts dominating above 25 keV. The spectra were also rebinned using the standard \texttt{ftgrouppha} task, with the optimal binning algorithm by \cite{Kaastra.Bleeker.2016} plus a minimum S/N of 3 per grouped bin. Type-I X-ray bursts were not detected in the light curves. No quasi-periodic oscillations were identified with the \nustar data. During the joint spectropolarimetric analysis, we only considered the \nustar data that were simultaneous with the \ixpe observation.

\subsection{\nicer}

The Neutron Star Interior Composition Explorer (\nicer; \citealt{Gendreau.etAl.2016}) observed \source from 2024 August 16 15:35:51 UT to August 17 00:56:40 UT in different snapshots for a total exposure time of 1.2 ks (see Table \ref{table:Obs}). The calibrated and cleaned files were processed using the standard \texttt{nicerl2} task of the \nicer Data Analysis Software (\textsc{nicerdas} v.13) together with the latest calibration files (CALDB v.20240206). The spectra and light curves were then obtained with the \texttt{nicerl3-spec} and \texttt{nicerl3-lc} commands, while the background was computed using the SCORPEON\footnote{\url{https://heasarc.gsfc.nasa.gov/docs/nicer/analysis_threads/scorpeon-overview}} model. Due to their short duration, it is not possible to build a complete CCD or HID from the \nicer observations. Neither Type-I X-ray bursts nor quasi-periodic oscillations were identified in the \nicer data. The \nicer data were considered in the 1--10 keV band. During spectropolarimetric analysis (Sect. \ref{sec:Data.Analysis}), we found some significant residuals in the \nicer energy spectra below 2--2.5 keV \citep{Miller.etAl.2018,Strohmayer.etAl.2018}, due to spectral features not included in the \nicer ancillary response file (ARF). We add a multiplicative absorption edge (\texttt{edge}) with threshold energies of $1.81 \pm 0.02$ and a Gaussian component (\texttt{gaussian}) at 1.7 keV with a line width of $0.06 \pm 0.01$ keV. 

%%%%%%%%%%%%%%%%% SPECTRA %%%%%%%%%%%%%%%%%%

\section{Data analysis}\label{sec:Data.Analysis}

\begin{figure}
    \centering
    \includegraphics[width=0.425\textwidth]{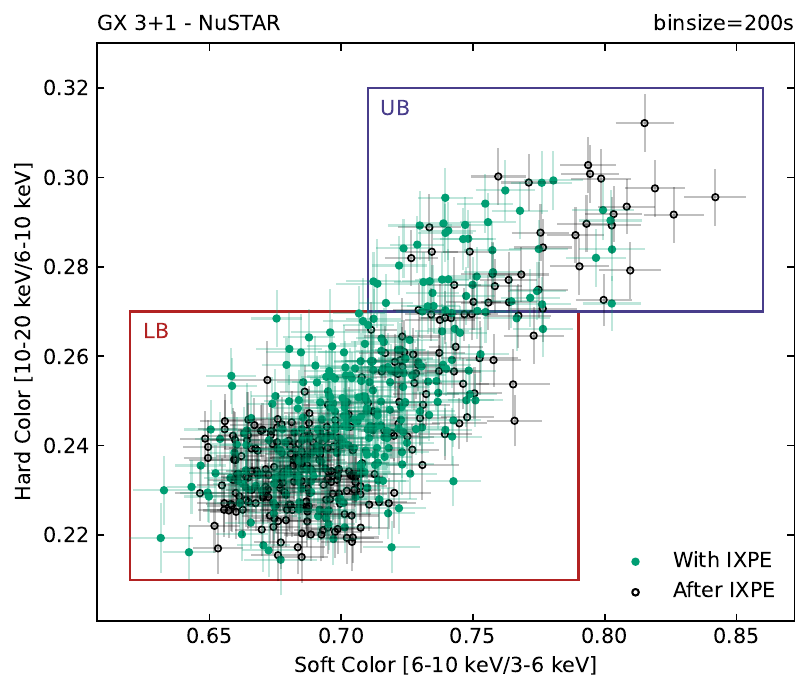}
    \caption{\nustar CCD of \source. The soft and hard colors are defined as the ratio of the counts in the 6--10/3--6 keV and 10--20/6--10 keV bands, respectively. Empty black circles denote the \nustar points not simultaneous with \ixpe observation. The red and purple boxes correspond to the LB and UB region. Each bin corresponds to 200 s.}
    \label{fig:CCD.Nustar}
\end{figure}

\subsection{Timing properties}

The light curves obtained with \ixpe, \nustar, and \nicer are shown in Fig. \ref{fig:LC}, using time bins of 200 s for \ixpe and \nustar, and 50 s for \nicer. Some significant variations in the observed flux can be noticed for all three observatories, with small drops in the observed count rate. Despite the fact that the \ixpe hardness ratio (HR: 5--8 keV/3--5 keV) remains quite constant throughout the observation, both \nustar and \nicer data exhibit some variability: in particular, \source moves along the CCD (Fig. \ref{fig:CCD.Nustar}) tracing part of the banana branch. The soft color is defined as the ratio of the counts in the 6--10/3--6 keV band, whereas the hard color as the ratio in the 10--20/6--10 keV energy range. % Soft and hard colors are defined as follows
% \begin{equation}
%     \label{eq:Soft.Col}
%     \text{Soft Color} = \frac{\text{Count rate}~[6-10~\mathrm{keV}]}{\text{Count rate}~[3-6~\mathrm{keV}]}~,
% \end{equation}
% \begin{equation}
%     \label{eq:Hard.Col}
%     \text{Hard Color} = \frac{\text{Count rate}~[10-20~\mathrm{keV}]}{\text{Count rate}~[6-10~\mathrm{keV}]}~.
% \end{equation}
Therefore, we divided the \nustar observation based on its hard color: when the hard color is less than 0.27, the source remains in the lower part of the CCD in the lower banana state (LB; red box in Fig. \ref{fig:CCD.Nustar}), while, when both the hard and soft colors increase, the source moves toward the upper banana branch (UB; purple box in Fig. \ref{fig:CCD.Nustar}). The gray regions in Fig. \ref{fig:LC} correspond to the time intervals during which the source moves along the UB branch. The same time intervals were then used to extract and filter in time for the \ixpe and \nicer observations.  

\subsection{Spectral analysis}

\begin{figure}
    \centering
    \includegraphics[width=0.465\textwidth]{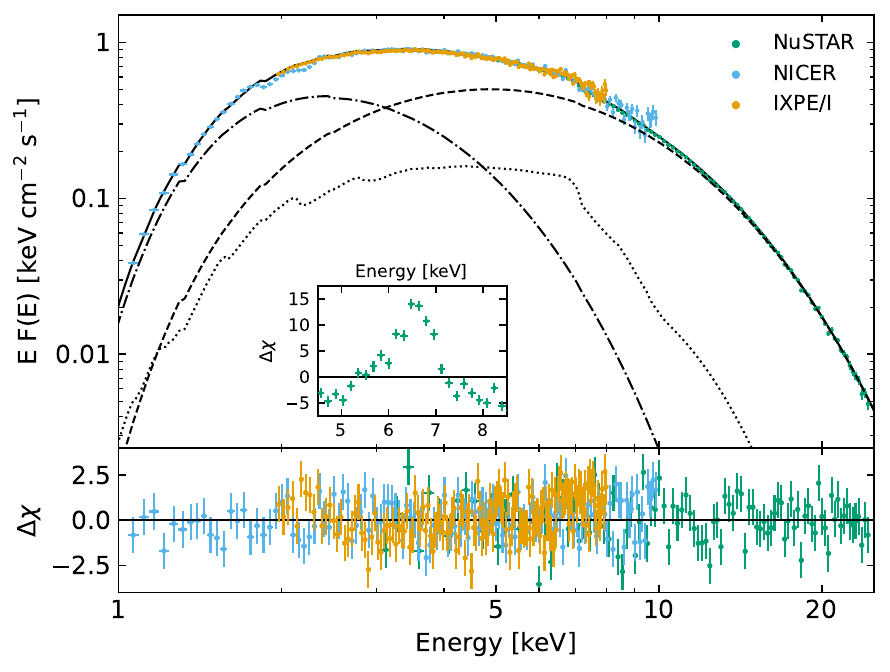}
    \includegraphics[width=0.475\textwidth]{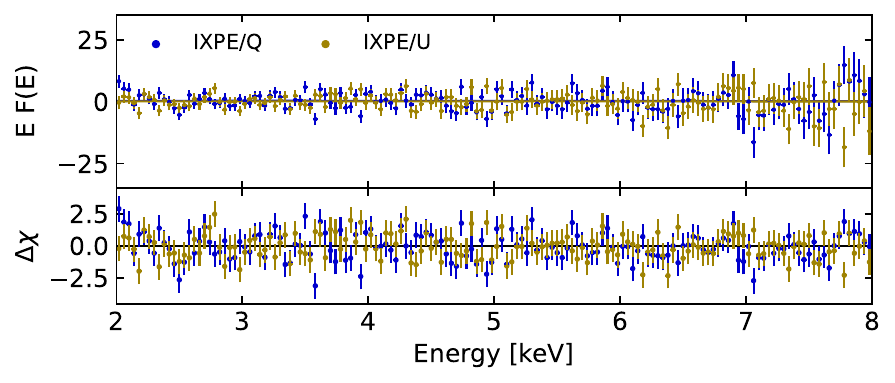}
    \caption{Deconvolved \ixpe (2--8 keV), \nustar (3--25 keV), \nicer (1--10 keV) spectra (top panels), and \ixpe $Q$ and $U$ Stokes spectra (bottom panels), with the best-fit model during the LB state and residuals in units of $\sigma$. We added the residuals without including the reflection components in the zoomed plot, highlighting the Fe K$\alpha$ line profile. The best-fit model includes \diskbb (dash-dotted lines), \texttt{thcomp*bbodyrad} (dashed lines), and \relxillns (dotted lines).}
    \label{fig:Spectra.LB}
\end{figure}
\begin{figure}
    \centering
    \includegraphics[width=0.465\textwidth]{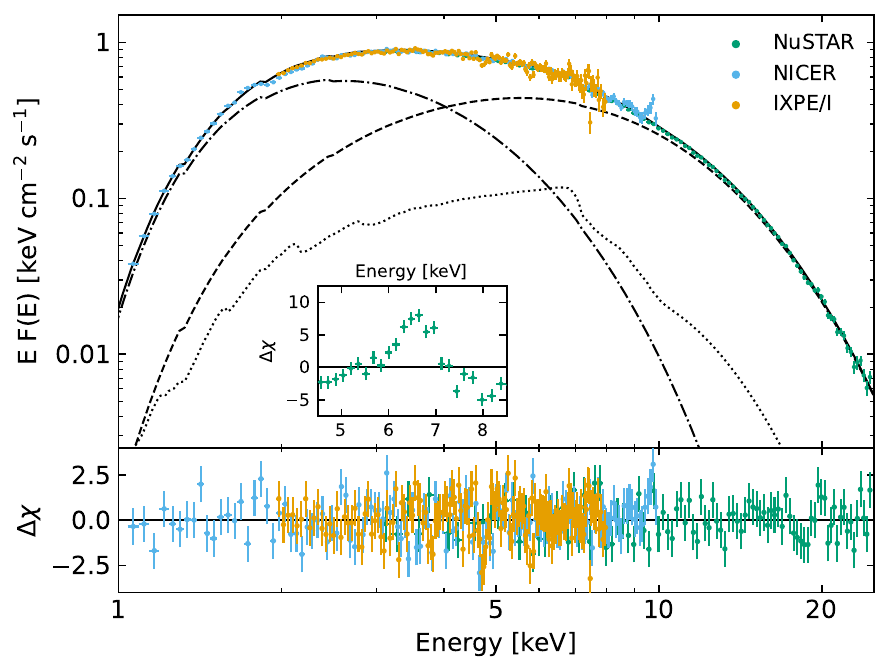}
    \includegraphics[width=0.475\textwidth]{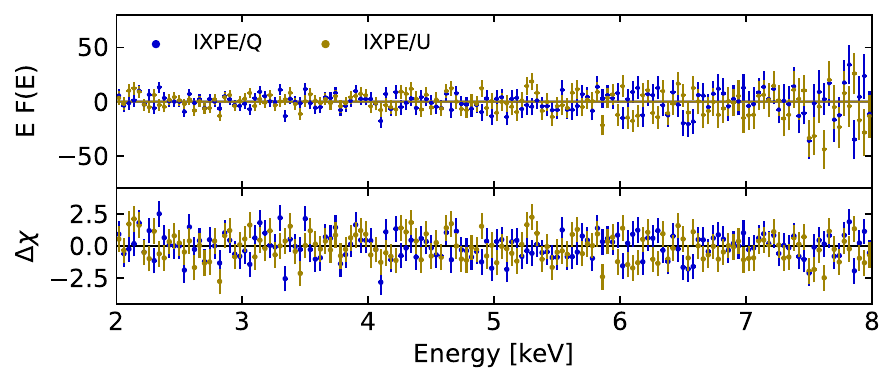}
    \caption{Deconvolved \ixpe (2--8 keV), \nustar (3--25 keV), \nicer (1--10 keV) spectra (top panels), and \ixpe $Q$ and $U$ Stokes spectra (bottom panels), with the best-fit model during the UB state and residuals in units of $\sigma$. We added the residuals without including the reflection components in the zoomed plot, highlighting the Fe K$\alpha$ line profile. The best-fit model includes \diskbb (dash-dotted lines), \texttt{thcomp*bbodyrad} (dashed lines), and \relxillns (dotted lines).}
    \label{fig:Spectra.UB}
\end{figure}

We performed a joint fit of the \ixpe, \nicer, and \nustar spectra during the LB and UB states with \textsc{xspec} \citep{Arnaud.1996}. We included energy-independent cross-calibration multiplicative factors for each \ixpe DU, each \nustar FPM, and the \nicer spectra. Then, we included the \texttt{TBabs} model with \texttt{vern} cross-section \citep{Verner.etAl.1996} and \texttt{wilm} abundances \citep{Wilms.etAl.2000} to take into account interstellar absorption, leaving the hydrogen column density $N_\text{H}$ free to vary. The best-fit value obtained ($2.41\times10^{22}{\rm\,cm^{-2}}$) is consistent with that obtained by \cite{Ludlam.etAl.2019}. The primary continuum is modeled with two components: a disk multi-temperature blackbody (\diskbb; \citealt{Mitsuda.etAl.1984}) dominating at lower energies plus a Comptonized blackbody emission using the convolution model \thcomp \citep{Zdziarski.etAl.2020} applied to the \bbodyrad component. The covering factor $f$ of \thcomp is related to the fraction of Comptonized photons and results to be $> 0.95$, with best-fit value of 0.99. Therefore, we fixed it to its best-fit value to improve the fit statistic and better constrain the other parameters of \thcomp. The physical parameters of \diskbb and \texttt{thcomp*bbodyrad} are considered separately between the two spectral states. We noticed some residuals in the \nicer and \nustar data, suggesting the presence of a strong iron K$\alpha$ line (see the zoomed plot in Figs. \ref{fig:Spectra.LB} and \ref{fig:Spectra.UB}) and a reflection component. Therefore, we added the reflection component to the model using \relxillns. The \texttt{relxill} models reproduce the relativistic reflection component from the innermost regions of the accretion disk \citep{Garcia.etAl.2014,Dauser.etAl.2014}. In particular, \relxillns considers a single-temperature blackbody spectrum that illuminates the surface of the accretion disk at 45\degr, which could physically originate from the emission of the NS surface or the spreading layer. The temperature of the seed photons of \relxillns is linked to the black body temperature of \bbodyrad of each spectral state. We fixed the number density at $\log n_e$/cm$^{-3}$ = 16.5, since the fit is not very sensitive to this parameter. This value is consistent with the results found by \cite{Ludlam.etAl.2019} and with the inner region of standard accretion disks \citep{Garcia.etAl.2016,Ludlam.etAl.2022}. Moreover, we also fixed the emissivity index $q_\text{em} = 2.8$ to the same value obtained by \cite{Ludlam.etAl.2019}, and the outer disk radius $R_\text{out} = 1000$ $R_\text{g}$, as the fit is not able to constrain these parameters. For a standard NS, the dimensionless spin $a$ can be derived using the spin period $P$ as $a = 0.47/P(\mathrm{ms})$ \citep{Braje.etAl.2000}. In particular, typical NSs in LMXBs are characterized by spin periods between 2 and 5 ms \citep{Patruno.2017,DiSalvo.etAl.2023}; therefore, the dimensionless spin can be fixed at $a=0.1$. We left the inclination of the system, the inner disk radius (in units of the ISCO), the ionization parameter $\xi$, the iron abundance $A_\text{Fe}$ and the normalization $N_{\rm r}$ free to vary. Since the best-fit values of all these parameters were very similar between the LB and UB spectra, we linked all the \relxillns parameters of the two spectral states to improve constraints on the results and we left only the two normalizations $N_{\rm r}$ separate.

\renewcommand{\arraystretch}{1.15}

\begin{table}
\caption{Best-fitting model parameters of the fits to \source \ixpe+\nustar+\nicer spectra.} 
\label{table:BestFit}      
\centering                                     
\begin{tabular}{l l c c}        
\hline\hline         
\noalign{\smallskip}
 & Parameter & LB & UB \\   
\hline     
\noalign{\smallskip}
\texttt{TBabs} & $N_{\rm H}$ ($10^{22}$\,cm$^{-2}$) & \multicolumn{2}{c}{2.41$^{+0.02}_{-0.02}$} \\
\texttt{diskbb} & $kT_\text{in}$ (keV) & 0.94$^{+0.05}_{-0.04}$ & 1.11$^{+0.04}_{-0.03}$ \\
 & $R_\text{in} \sqrt{\cos i}$ (km) & 11.4$^{+1.7}_{-1.5}$ & 9.3$^{+1.6}_{-1.5}$ \\
 \texttt{thcomp} & $\tau$ & 7.7$^{+0.2}_{-0.2}$ & 7.9$^{+1.9}_{-2.1}$ \\
 & $kT_{\rm e}$ (keV) & 2.8$^{+0.1}_{-0.1}$ & 2.7$^{+0.2}_{-0.2}$ \\
 % & $f$ & [1] & [1] \\
\texttt{bbodyrad} & $kT$ (keV) & 1.39$^{+0.04}_{-0.04}$ & 1.65$^{+0.11}_{-0.12}$ \\
 & $R_\text{bb}$ (km) & 7.9$^{+0.7}_{-0.7}$ & 5.6$^{+1.7}_{-0.7}$ \\
\texttt{relxillNS} & $q_\text{em}$ & \multicolumn{2}{c}{[2.8]} \\
 & $a$ & \multicolumn{2}{c}{[0.1]} \\
 & $i$ (deg) & \multicolumn{2}{c}{36.1$^{+1.1}_{-1.9}$} \\
 & $R_\text{in}$ (ISCO) & \multicolumn{2}{c}{< 1.5} \\
 % & $R_\text{out}$ (ISCO) & [1000] & [1000] \\
 & $kT_\text{bb}$ (keV) & = $kT$ & = $kT$ \\
 & $\log \xi$ & \multicolumn{2}{c}{2.8$^{+0.1}_{-0.1}$} \\
 & $A_\text{Fe}$ & \multicolumn{2}{c}{1.7$^{+0.1}_{-0.1}$} \\
 & $\log n_{\rm e}$ & \multicolumn{2}{c}{[16.5]} \\
 & $N_{\rm r}$ ($10^{-3}$) & 4.2$^{+0.5}_{-0.5}$ & 3.0$^{+0.4}_{-0.3}$ \\
\hline
\multicolumn{4}{c}{Cross-calibration} \\
  \multicolumn{2}{l}{$\mathcal{C}_\text{DU1/FPMA}$} & 0.864$^{+0.003}_{-0.003}$ & 0.843$^{+0.006}_{-0.006}$ \\
  \multicolumn{2}{l}{$\mathcal{C}_\text{DU2/FPMA}$} & 0.869$^{+0.003}_{-0.003}$ & 0.848$^{+0.006}_{-0.006}$ \\
  \multicolumn{2}{l}{$\mathcal{C}_\text{DU3/FPMA}$} & 0.849$^{+0.003}_{-0.003}$ & 0.827$^{+0.006}_{-0.006}$ \\
  \multicolumn{2}{l}{$\mathcal{C}_\text{FPMB/FPMA}$} & 0.982$^{+0.002}_{-0.002}$ & 0.978$^{+0.002}_{-0.002}$ \\
  \multicolumn{2}{l}{$\mathcal{C}_\text{XTI/FPMA}$} & 0.933$^{+0.002}_{-0.003}$ & 0.861$^{+0.003}_{-0.003}$ \\
\hline
  \multicolumn{2}{l}{$\chi^2/\text{d.o.f.}$} & 880.9/817 & 812.3/800 \\
\hline
\multicolumn{4}{c}{Photon flux ratios (2--8 keV)} \\
  \multicolumn{2}{l}{$F_\texttt{diskbb}/F_\text{Tot}$} & 34.3\% & 40.1\% \\
  \multicolumn{2}{l}{$F_\texttt{thcomp*bbodyrad}/F_\text{Tot}$} & 47.9\% & 47.8\% \\
  \multicolumn{2}{l}{$F_\texttt{relxillNS}/F_\text{Tot}$} & 16.8\% & 12.1\% \\
\multicolumn{4}{c}{Energy Flux (2--8 keV)} \\
  \multicolumn{2}{l}{$F_\text{Tot}$ ($10^{-9}$ \fluxcgs)} & 7.03 & 7.05 \\
\hline
\end{tabular}
\tablefoot{
Uncertainties are at the 90$\%$ confidence level for a single parameter. Parameters in common between the two columns were linked during the two spectral states, while those in square brackets were frozen during the fit. The normalizations of \diskbb and \bbodyrad are computed assuming a source distance of 6.5 kpc \citep{Galloway.etAl.2008}. The cross-calibration constant are reported with respect to that of the \nustar FPMA detector.}
\end{table}

The best-fit is obtained considering the following spectral model: 
\begin{description}
\item \texttt{TBabs*(diskbb + thcomp*bbodyrad + relxillNS)} ~,
\end{description}
which is capable of providing excellent fits to the \ixpe+\nustar+\nicer spectra. The best-fit values for each parameter are reported in Table \ref{table:BestFit}, while the spectra are represented in Figs. \ref{fig:Spectra.LB} and \ref{fig:Spectra.UB}. The physical properties of both the disk and Comptonized components can be well constrained for both spectral states: the \diskbb and \bbodyrad temperature increase as the source moves from the LB to the UB state, while the best-fit values for the electron temperature $kT_{\rm e}$ and optical depth $\tau$ of the Comptonizing region remain quite stable, differently from other Atoll sources, such as 4U~1728--34 \citep{Tarana.etAl.2011}. The inclination $i$ of the system can be well constrained from the reflection component, suggesting that the inclination is about 36\degr, while only an upper limit of 1.4 $R_\text{ISCO}$ is obtained for the inner edge of the accretion disk. The ionization parameter $\xi$ and the iron abundance $A_\text{Fe}$ can also be obtained from the best-fit model with good precision.

The parameters obtained from the best-fit models are typical for bright atoll NS-LMXBs in the soft state. The apparent inner disk radius $R_\text{in}$ and the radius of the black body photon-emitting region $R_\text{bb}$ can be computed from the normalizations of \diskbb and \bbodyrad assuming a 6.5 kpc distance \citep{Galloway.etAl.2008} and that all seed photons of \bbodyrad are Comptonized, which corresponds to $f=1$. We found that the black body-emitting region has an equivalent spherical radius of $\sim 8$ km during the LB state and it decreases up to $\sim 5.5$ km during the UB intervals, consistent with seed photons originating in a region consistent with the spreading layer. From \diskbb, the inner disk radius results to be close to $\sim 10 \sqrt{\cos i}$ km, which is consistent with the upper limit derived from the reflection component using \relxillns. 

\subsection{Polarimetric analysis}

\renewcommand{\arraystretch}{1.05}

\begin{figure}
    \centering
    \includegraphics[width=0.425\textwidth]{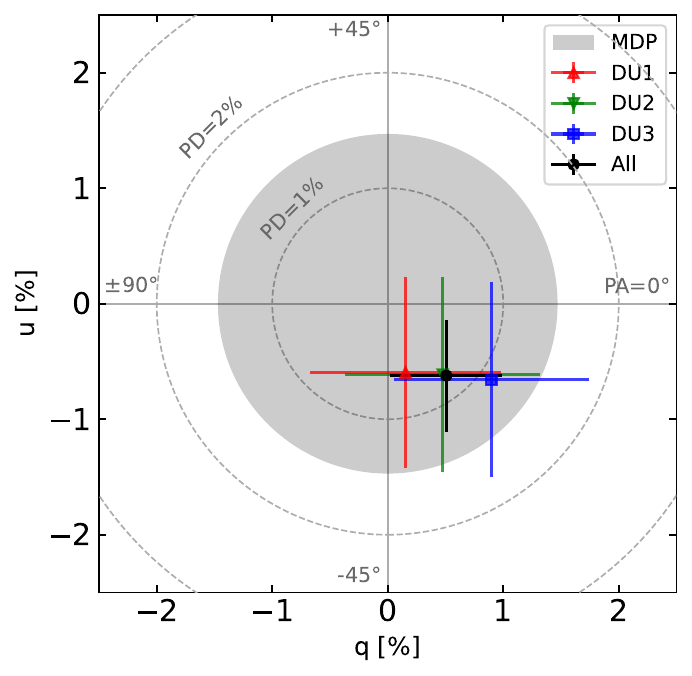}
    \caption{Normalized Stokes $q$ ($Q/I$) and $u$ ($U/I$) parameters in the 2--8 keV band obtained with the \texttt{PCUBE} algorithm of \textsc{ixpeobssim} \citep{Baldini.etAl.2022}, for the three \ixpe DUs and their combination. The gray-filled circle corresponds to the 99\% MDP.}
    \label{fig:Stokes}
\end{figure}
\begin{table}
\caption{Polarization measured with \textsc{xspec}.}             
\label{table:Pol}      
\centering                                     
\begin{tabular}{l c}         
\hline\hline       
  Energy range & PD Upper Limit \\   
\hline     
  2--8 keV & $<1.3\%$ \\
  2--4 keV & $<1.7\%$ \\
  4--8 keV & $<1.8\%$ \\
\hline                                            
\end{tabular}
\tablefoot{
Upper limits are reported at the 99\% confidence level.
}
\end{table}

\begin{figure}
    \centering
    \includegraphics[width=0.425\textwidth]{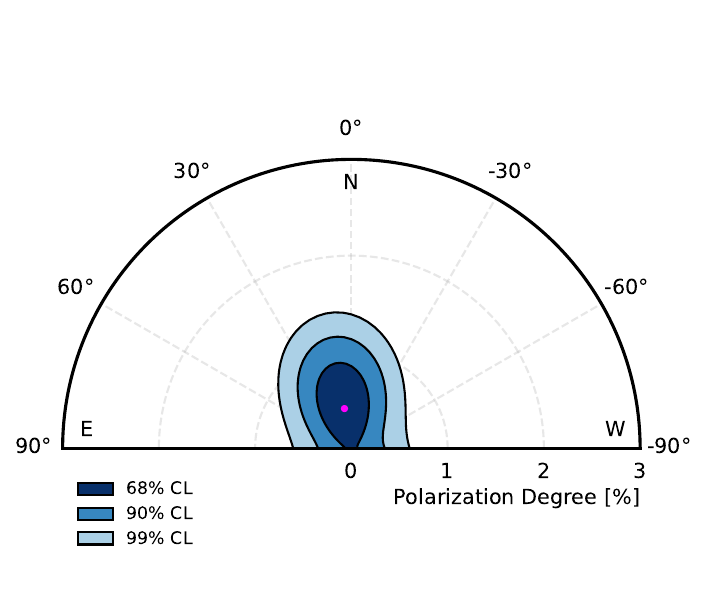}
    \caption{Contour plots of the PD and PA at the 68\%, 90\%, and 99\% confidence levels, in the 2--8 keV energy bands obtained with \textsc{xspec}.}
    \label{fig:Contours}
\end{figure}

To estimate polarization in the 2--8 keV band of \source, we first performed a polarimetric analysis of the entire \ixpe observation using \textsc{ixpeobssim} (v.31.0.1; \citealt{Baldini.etAl.2022}), with the latest available calibration file (v.13.20240701). The normalized Stokes parameters measured by \ixpe obtained using the \texttt{PCUBE} task of \textsc{ixpeobssim} are shown in Fig. \ref{fig:Stokes} for each DU, with the minimum detectable polarization (MDP) at the 99\% level. The spectropolarimetric fitting procedure is performed with \textsc{xspec} by applying \polconst to the best-fit models.
% \begin{description}
% \item \texttt{polconst*TBabs*(diskbb + thcomp*bbodyrad + relxillNS)}~.
% \end{description}
No significant detection is obtained in the 2--8 keV band, with an upper limit to the polarization degree of 1.3\% at the 99\% confidence level for a single parameter of interest, which is independent of the value of the polarization angle. In Table \ref{table:Pol} we report the polarization degree and angle measured in different energy bins obtained with \textsc{xspec} \citep{Arnaud.1996}, and the corresponding upper limits. The two-dimensional polarization contours in the full 2--8 keV band obtained with \textsc{xspec} are represented in Fig. \ref{fig:Contours}. We tried also to separate the full \ixpe band considering energy bins of 1 or 2 keV, but without any significant detection of the polarization in the considered energy range. Therefore, we did not find any evident trend of the polarization with energy. We also analyzed the polarimetric data for each GTI: during the time intervals with lower hard color, the polarization degree is <1.4\%, while when the soft and hard color increases, the upper limit of the polarization degree is 2.3\%. Even by separating the data by hard color, we were not able to detect any significant evolution with energy. All results are consistent between the \texttt{PCUBE} task and the fitting procedure with \textsc{xspec}.

%%%%%%%%%%%%%%%%% RESULTS %%%%%%%%%%%%%%%%%%

\section{Discussion}\label{sec:Discussion}

The systematic effects on the derived polarization are expected to be minimized by the overlap of the energy range between \ixpe and \nustar+\nicer. Due to the limited band-pass of \ixpe and the undetected polarization in the 2--8 keV energy band, it is difficult to estimate the polarization degree and angle for each spectral component. In particular, the Comptonization and reflection components exhibit similar spectral shapes, also peaking at similar energies. Therefore, we need some theoretical assumptions to find an estimate of the polarization of the different components.

\begin{table}
\caption{Polarization degree and angle of each spectral component for different scenarios.} 
\label{table:Pol.Comp}      
\centering                                     
\begin{tabular}{lll}        
\hline\hline         
\noalign{\smallskip}
 Component & PD (\%) & PA (deg) \\   
\hline     
\noalign{\smallskip}
\diskbb & $1.7 \pm 1.4$ & $4 \pm 27$ \\
\texttt{thcomp*bbodyrad} & [0] & - \\
\relxillns & [0] & - \\
\hline
\diskbb & [0] & - \\
\texttt{thcomp*bbodyrad} & $<2.6$ & Unconstrained \\
\relxillns & [0] & - \\
\hline
\diskbb & [0] & - \\
\texttt{thcomp*bbodyrad} & [0] & - \\
\relxillns & $<8.1$ & Unconstrained \\
\hline
\diskbb & [0.5] & = $\text{PA}_\text{Refl} - 90\degr$ \\
\texttt{thcomp*bbodyrad} & [0.5] & = $\text{PA}_\text{Refl}$ \\
\relxillns & $<7.8$ & Unconstrained \\
\hline
\diskbb & [0.5] & = $\text{PA}_\text{Comp} - 90\degr$ \\
\texttt{thcomp*bbodyrad} & $<0.7$ & Unconstrained \\
\relxillns & [10] & = $\text{PA}_\text{Comp}$ \\
\hline
\end{tabular}
\tablefoot{
The errors are at the 90\% confidence level, while upper limits are reported at 99\% confidence level for one interesting parameter. Parameters in square brackets are frozen.
}
\end{table}

We performed the spectropolarimetric fitting procedure by applying \polconst to each spectral component: 
\begin{description}
\item {\tt TBabs*(polconst}$^{(d)}${\tt*diskbb + polconst}$^{(r)}${\tt*relxillNS + polconst}$^{(c)}${\tt *thcomp*bbodyrad)}~.
\end{description}
During the analysis, we fixed all spectral parameters to their best-fit values (see Table \ref{table:BestFit}), except for the cross-calibration constants, and left only the \polconst polarization degree and angle free to vary. The results are reported in Table \ref{table:Pol.Comp}. As a first attempt, we tried to estimate the polarization of each spectral component assuming that only one is polarized and the other two have zero polarization (see Table \ref{table:Pol.Comp}). The only component for which we were able to constrain (at the 90\% confidence level) the polarimetric contribution is \diskbb, with $1.7\% \pm 1.4\%$ PD, while only upper limits were obtained for Comptonization ($<2.6\%$, at the 99\% confidence level) and reflection ($<8.1\%$, at the 99\% confidence level). In all three cases, the fit results were acceptable with $\chi^2/\text{d.o.f.} = 1.05$. Although the results are perfectly consistent with the theoretical predictions for a semi-infinite plane-parallel atmosphere \citep{Chandrasekhar.1960} for the disk component and with a typical spreading layer geometry for Comptonization \citep{Gnarini.etAl.2022,Ursini.etAl.2023,Farinelli.etAl.2024,Bobrikova.etAl.2024.SL}, the upper limit on the reflection polarization seems to be rather stringent. Reflected photons are expected to be highly polarized \citep{Matt.1993,Poutanen.etAl.1996} and can contribute significantly to the polarization signal even if their fraction with respect to the total flux is relatively low. For \source, the contribution of reflection to the total flux in the 2--8 keV energy range is quite high ($\approx 15\%$). The resulting upper limit is slightly lower than the theoretical previsions for reflected radiation. We also tried to fix the disk PD to the classical value obtained by \cite{Chandrasekhar.1960} and the PD of the Comptonization component to 0.5\% (see Table \ref{table:Pol.Comp}), a reasonable value for a spreading layer configuration observed at 36\degr. The PA of \diskbb is fixed to be orthogonal to that of the reflected and Comptonized photons. With these assumptions, the fit does not improve while the upper limit slightly decreases ($<7.8\%$ at the 99\% confidence level). We also tried a different scenario in which the reflection is highly polarized as expected, with PD fixed at 10\% \citep{Matt.1993}, while the polarization of \texttt{thcomp*bbodyrad} is left free to vary. The PD of \diskbb is set again to the results obtained by \cite{Chandrasekhar.1960} for $i \approx 36$\degr, polarized orthogonally with respect to the other two components. We obtained only an upper limit of 0.7\% (at the 99\% confidence level) for the PD of the Comptonized component, suggesting a spherical or a spreading layer-like geometry of the Comptonizing region \citep{Gnarini.etAl.2022,Farinelli.etAl.2024,Bobrikova.etAl.2024.SL}. 

%%%%%%%%%%%%%%%%% CONCLUSIONS %%%%%%%%%%%%%%%%%%

\section{Conclusions}\label{sec:Conclusions}

We report the results of the first spectropolarimetric observation of the bright atoll NS-LMXB \source performed with \ixpe, with joint \nicer and \nustar observations. Only an upper limit of 1.3\% (at the 99\% confidence level) can be derived from the \ixpe data. The source was found in the banana branch, mainly on the lower part but moving towards the upper branch for short time intervals. The spectral properties remain quite stable during the motion along the banana branch: the spectra are well fitted with a two-component model, that is, disk plus Comptonized emission, with the addition of the reflected photons to remove some residuals at high energies and in the 6--7 keV energy range. From the reflection component, we were able to obtain an upper limit on the position of the inner edge of the accretion disk ($R_\text{in} \lesssim 1.5$ $R_\text{ISCO}$) and to estimate the inclination of the system ($i \approx 36$\degr). Therefore, the low polarization signal can be mainly due to the inclination, since more face-on systems are expected to be less polarized \citep{Chandrasekhar.1960,Gnarini.etAl.2022,Farinelli.etAl.2024}. Because we obtained only an upper limit on the polarization, we were not able to constrain the polarimetric properties of the different spectral components. However, we found a quite stringent upper limit for the PD of the reflection component: differently from other NS-LMXBs observed by \ixpe, considering the observed photon flux ratios derived from the spectra, the reflected photons are expected to be less polarized (up to $\approx 8\%$); otherwise we would have obtained a polarization detection if not in the entire 2-8 keV band at least in some energy ranges. The upper limits on the disk and Comptonization component are still consistent with the theoretical expectation for a system with a standard accretion disk and a spreading layer-like geometry of the Comptonizing region.

%--------------------------------------------------------------------

\begin{acknowledgements}
The Imaging X-ray Polarimetry Explorer (IXPE) is a joint US and Italian mission. The US contribution is supported by the National Aeronautics and Space Administration (NASA) and led and managed by its Marshall Space Flight Center (MSFC). AG, SB, FC, SF, GM, AT, and FU acknowledge financial support by the Italian Space Agency (Agenzia Spaziale Italiana, ASI) through the contract ASI-INAF-2022-19-HH.0.
This research was also supported by the Istituto Nazionale di Astrofisica (INAF) grant 1.05.23.05.06: ``Spin and Geometry in accreting X-ray binaries: The first multi frequency spectro-polarimetric campaign''. 
SF has been supported by the project PRIN 2022 - 2022LWPEXW - ``An X-ray view of compact objects in polarized light'', CUP C53D23001180006.
AB acknowledges support from the Finnish Cultural Foundation grant 00240328. 
JP thanks the Academy of Finland grant 333112 for support. M.N. is a Fonds de Recherche du Quebec – Nature et Technologies (FRQNT) postdoctoral fellow.

This research used data products and software provided by the \ixpe, the \nicer and \nustar teams and distributed with additional software tools by the High-Energy Astrophysics Science Archive Research Center (HEASARC), at NASA Goddard Space Flight Center (GSFC). %This work was supported in part by NASA through the \nicer mission and the Astrophysics Explorers Program, together with the \nustar mission, a project led by the California Institute of Technology, managed by the Jet Propulsion Laboratory, and funded by the National Aeronautics and Space Administration. The \nustar Data Analysis Software (\textsc{NUSTARDAS}), jointly developed by the ASI Science Data Center (ASDC, Italy) and the California Institute of Technology (USA), has also been used in this project. 
\end{acknowledgements}


\bibliographystyle{aa} 
\bibliography{Bibliography} 

\begin{thebibliography}{78}
\expandafter\ifx\csname natexlab\endcsname\relax\def\natexlab#1{#1}\fi

\bibitem[{{Arnaud}(1996)}]{Arnaud.1996}
{Arnaud}, K.~A. 1996, in ASP Conf. Ser., Vol. 101, Astronomical Data Analysis
  Software and Systems V, ed. G.~H. {Jacoby} \& J.~{Barnes} (San Francisco:
  Astron. Soc. Pac.), 17--20

\bibitem[{{Asai} {et~al.}(1993){Asai}, {Dotani}, {Nagase}, {Mitsuda},
  {Kitamoto}, {Makishima}, {Takeshima}, \& {Kawabata}}]{Asai.etAl.1993}
{Asai}, K., {Dotani}, T., {Nagase}, F., {et~al.} 1993, \pasj, 45, 801

\bibitem[{{Baldini} {et~al.}(2022){Baldini}, {Bucciantini}, {Lalla}, {Ehlert},
  {Manfreda}, {Negro}, {Omodei}, {Pesce-Rollins}, {Sgr{\`o}}, \&
  {Silvestri}}]{Baldini.etAl.2022}
{Baldini}, L., {Bucciantini}, N., {Lalla}, N.~D., {et~al.} 2022, SoftwareX, 19,
  101194

\bibitem[{{Bobrikova} {et~al.}(2024){Bobrikova}, {Poutanen}, \&
  {Loktev}}]{Bobrikova.etAl.2024.SL}
{Bobrikova}, A., {Poutanen}, J., \& {Loktev}, V. 2024, \aap, submitted,
  arXiv:2409.16023

\bibitem[{{Bowyer} {et~al.}(1965){Bowyer}, {Byram}, {Chubb}, \&
  {Friedman}}]{Bowyer.etAl.1965}
{Bowyer}, S., {Byram}, E.~T., {Chubb}, T.~A., \& {Friedman}, H. 1965, Science,
  147, 394

\bibitem[{{Braje} {et~al.}(2000){Braje}, {Romani}, \&
  {Rauch}}]{Braje.etAl.2000}
{Braje}, T.~M., {Romani}, R.~W., \& {Rauch}, K.~P. 2000, \apj, 531, 447

\bibitem[{{Capitanio} {et~al.}(2023){Capitanio}, {Fabiani}, {Gnarini},
  {Ursini}, {Ferrigno}, {Matt}, {Poutanen}, {Cocchi}, {Mikusincova},
  {Farinelli}, {Bianchi}, {Kajava}, {Muleri}, {Sanchez-Fernandez}, {Soffitta},
  {Wu}, {Agudo}, {Antonelli}, {Bachetti}, {Baldini}, {Baumgartner},
  {Bellazzini}, {Bongiorno}, {Bonino}, {Brez}, {Bucciantini}, {Castellano},
  {Cavazzuti}, {Ciprini}, {Costa}, {De Rosa}, {Del Monte}, {Di Gesu}, {Di
  Lalla}, {Di Marco}, {Donnarumma}, {Doroshenko}, {Dov{\v{c}}iak}, {Ehlert},
  {Enoto}, {Evangelista}, {Ferrazzoli}, {Garcia}, {Gunji}, {Hayashida}, {Heyl},
  {Iwakiri}, {Jorstad}, {Karas}, {Kitaguchi}, {Kolodziejczak}, {Krawczynski},
  {La Monaca}, {Latronico}, {Liodakis}, {Maldera}, {Manfreda}, {Marin},
  {Marinucci}, {Marscher}, {Marshall}, {Mitsuishi}, {Mizuno}, {Ng}, {O'Dell},
  {Omodei}, {Oppedisano}, {Papitto}, {Pavlov}, {Peirson}, {Perri},
  {Pesce-Rollins}, {Petrucci}, {Pilia}, {Possenti}, {Puccetti}, {Ramsey},
  {Rankin}, {Ratheesh}, {Romani}, {Sgr{\`o}}, {Slane}, {Spandre}, {Tamagawa},
  {Tavecchio}, {Taverna}, {Tawara}, {Tennant}, {Thomas}, {Tombesi}, {Trois},
  {Tsygankov}, {Turolla}, {Vink}, {Weisskopf}, {Xie}, \&
  {Zane}}]{Capitanio.etAl.2023}
{Capitanio}, F., {Fabiani}, S., {Gnarini}, A., {et~al.} 2023, \apj, 943, 129

\bibitem[{{Chandrasekhar}(1960)}]{Chandrasekhar.1960}
{Chandrasekhar}, S. 1960, {Radiative transfer} (New York: Dover Publications)

\bibitem[{{Chenevez} {et~al.}(2006){Chenevez}, {Falanga}, {Brandt},
  {Farinelli}, {Frontera}, {Goldwurm}, {in't Zand}, {Kuulkers}, \&
  {Lund}}]{Chenevez.etAl.2006}
{Chenevez}, J., {Falanga}, M., {Brandt}, S., {et~al.} 2006, \aap, 449, L5

\bibitem[{{Cocchi} {et~al.}(2023){Cocchi}, {Gnarini}, {Fabiani}, F., {Poutanen,
  Juri}, {Capitanio, Fiamma}, {Bobrikova, Anna}, {Farinelli, Ruben}, {Paizis,
  Adamantia}, {Sidoli, Lara}, {Veledina, Alexandra}, {Bianchi, Stefano}, {Di
  Marco, Alessandro}, {Ingram, Adam}, {Kajava, Jari J. E.}, {La Monaca, Fabio},
  {Matt, Giorgio}, {Malacaria, Christian}, {Mikusincov\'a, Romana}, {Rankin,
  John}, {Zane, Silvia}, {Agudo, Iv\'an}, {Antonelli, Lucio A.}, {Bachetti,
  Matteo}, {Baldini, Luca}, {Baumgartner, Wayne H.}, {Bellazzini, Ronaldo},
  {Bongiorno, Stephen D.}, {Bonino, Raffaella}, {Brez, Alessandro},
  {Bucciantini, Niccol\`o}, {Castellano, Simone}, {Cavazzuti, Elisabetta},
  {Chen, Chien-Ting}, {Ciprini, Stefano}, {Costa, Enrico}, {De Rosa,
  Alessandra}, {Del Monte, Ettore}, {Di Gesu, Laura}, {Di Lalla, Niccol\`o},
  {Donnarumma, Immacolata}, {Doroshenko, Victor}, {Dovciak, Michal}, {Ehlert,
  Steven R.}, {Enoto, Teruaki}, {Evangelista, Yuri}, {Ferrazzoli, Riccardo},
  {Garcia, Javier A.}, {Gunji, Shuichi}, {Hayashida, Kiyoshi}, {Heyl, Jeremy},
  {Iwakiri, Wataru}, {Jorstad, Svetlana G.}, {Kaaret, Philip}, {Karas,
  Vladimir}, {Kislat, Fabian}, {Kitaguchi, Takao}, {Kolodziejczak, Jeffery J.},
  {Krawczynski, Henric}, {Latronico, Luca}, {Liodakis, Ioannis}, {Maldera,
  Simone}, {Manfreda, Alberto}, {Marin, Fr\'ed\'eric}, {Marinucci, Andrea},
  {Marscher, Alan P.}, {Marshall, Herman L.}, {Massaro, Francesco}, {Mitsuishi,
  Ikuyuki}, {Mizuno, Tsunefumi}, {Muleri, Fabio}, {Negro, Michela}, {Ng,
  Chi-Yung}, {O\'{}Dell, Stephen L.}, {Omodei, Nicola}, {Oppedisano, Chiara},
  {Papitto, Alessandro}, {Pavlov, George G.}, {Peirson, Abel L.}, {Perri,
  Matteo}, {Pesce-Rollins, Melissa}, {Petrucci, Pierre-Olivier}, {Pilia,
  Maura}, {Possenti, Andrea}, {Puccetti, Simonetta}, {Ramsey, Brian D.},
  {Ratheesh, Ajay}, {Roberts, Oliver J.}, {Romani, Roger W.}, {Sgr\`o,
  Carmelo}, {Slane, Patrick}, {Soffitta, Paolo}, {Spandre, Gloria}, {Swartz,
  Douglas A.}, {Tamagawa, Toru}, {Tavecchio, Fabrizio}, {Taverna, Roberto},
  {Tawara, Yuzuru}, {Tennant, Allyn F.}, {Thomas, Nicholas E.}, {Tombesi,
  Francesco}, {Trois, Alessio}, {Tsygankov, Sergey S.}, {Turolla, Roberto},
  {Vink, Jacco}, {Weisskopf, Martin C.}, {Wu, Kinwah}, \& {Xie,
  Fei}}]{Cocchi.etAl.2023}
{Cocchi}, M., {Gnarini}, A., {Fabiani}, S., {et~al.} 2023, \aap, 674, L10

\bibitem[{{Costa} {et~al.}(2001){Costa}, {Soffitta}, {Bellazzini}, {Brez},
  {Lumb}, \& {Spandre}}]{Costa.2001}
{Costa}, E., {Soffitta}, P., {Bellazzini}, R., {et~al.} 2001, \nat, 411, 662

\bibitem[{{Dauser} {et~al.}(2014){Dauser}, {Garcia}, {Parker}, {Fabian}, \&
  {Wilms}}]{Dauser.etAl.2014}
{Dauser}, T., {Garcia}, J., {Parker}, M.~L., {Fabian}, A.~C., \& {Wilms}, J.
  2014, \mnras, 444, L100

\bibitem[{{den Hartog} {et~al.}(2003){den Hartog}, {in't Zand}, {Kuulkers},
  {Cornelisse}, {Heise}, {Bazzano}, {Cocchi}, {Natalucci}, \&
  {Ubertini}}]{denHartog.etAl.2003}
{den Hartog}, P.~R., {in't Zand}, J.~J.~M., {Kuulkers}, E., {et~al.} 2003,
  \aap, 400, 633

\bibitem[{{Di Marco} {et~al.}(2022){Di Marco}, {Costa}, {Muleri}, {Soffitta},
  {Fabiani}, {La Monaca}, {Rankin}, {Xie}, {Bachetti}, {Baldini},
  {Baumgartner}, {Bellazzini}, {Brez}, {Castellano}, {Del Monte}, {Di Lalla},
  {Ferrazzoli}, {Latronico}, {Maldera}, {Manfreda}, {O'Dell}, {Perri},
  {Pesce-Rollins}, {Puccetti}, {Ramsey}, {Ratheesh}, {Sgr{\`o}}, {Spandre},
  {Tennant}, {Tobia}, {Trois}, \& {Weisskopf}}]{DiMarco.etAl.2022}
{Di Marco}, A., {Costa}, E., {Muleri}, F., {et~al.} 2022, \aj, 163, 170

\bibitem[{{Di Marco} {et~al.}(2023{\natexlab{a}}){Di Marco}, {La Monaca},
  {Poutanen}, {Russell}, {Anitra}, {Farinelli}, {Mastroserio}, {Muleri}, {Xie},
  {Bachetti}, {Burderi}, {Carotenuto}, {Del Santo}, {Di Salvo},
  {Dov{\v{c}}iak}, {Gnarini}, {Iaria}, {Kajava}, {Liu}, {Middei}, {O'Dell},
  {Pilia}, {Rankin}, {Sanna}, {Eijnden}, {Weisskopf}, {Bobrikova}, {Capitanio},
  {Costa}, {Kaaret}, {Marino}, {Soffitta}, {Ursini}, {Ambrosino}, {Cocchi},
  {Fabiani}, {Marshall}, {Matt}, {Motta}, {Papitto}, {Stella}, {Tarana},
  {Zane}, {Agudo}, {Antonelli}, {Baldini}, {Baumgartner}, {Bellazzini},
  {Bianchi}, {Bongiorno}, {Bonino}, {Brez}, {Bucciantini}, {Castellano},
  {Cavazzuti}, {Chen}, {Ciprini}, {De Rosa}, {Del Monte}, {Di Gesu}, {Di
  Lalla}, {Donnarumma}, {Doroshenko}, {Ehlert}, {Enoto}, {Evangelista},
  {Ferrazzoli}, {Garcia}, {Gunji}, {Hayashida}, {Heyl}, {Iwakiri}, {Jorstad},
  {Karas}, {Kislat}, {Kitaguchi}, {Kolodziejczak}, {Krawczynski}, {Latronico},
  {Liodakis}, {Maldera}, {Manfreda}, {Marin}, {Marinucci}, {Marscher},
  {Massaro}, {Mitsuishi}, {Mizuno}, {Negro}, {Ng}, {Omodei}, {Oppedisano},
  {Pavlov}, {Peirson}, {Perri}, {Pesce-Rollins}, {Petrucci}, {Possenti},
  {Puccetti}, {Ramsey}, {Ratheesh}, {Roberts}, {Romani}, {Sgr{\`o}}, {Slane},
  {Spandre}, {Swartz}, {Tamagawa}, {Tavecchio}, {Taverna}, {Tawara}, {Tennant},
  {Thomas}, {Tombesi}, {Trois}, {Tsygankov}, {Turolla}, {Vink}, {Wu}, \& {IXPE
  Collaboration}}]{DiMarco.etAl.2023.4U}
{Di Marco}, A., {La Monaca}, F., {Poutanen}, J., {et~al.} 2023{\natexlab{a}},
  \apjl, 953, L22

\bibitem[{{Di Marco} {et~al.}(2023{\natexlab{b}}){Di Marco}, {Soffitta},
  {Costa}, {Ferrazzoli}, {La Monaca}, {Rankin}, {Ratheesh}, {Xie}, {Baldini},
  {Del Monte}, {Ehlert}, {Fabiani}, {Kim}, {Muleri}, {O'Dell}, {Ramsey},
  {Rubini}, {Sgr{\`o}}, {Silvestri}, {Tennant}, \&
  {Weisskopf}}]{DiMarco.etAl.2023}
{Di Marco}, A., {Soffitta}, P., {Costa}, E., {et~al.} 2023{\natexlab{b}}, \aj,
  165, 143

\bibitem[{{Di Salvo} {et~al.}(2009){Di Salvo}, {D'A{\'\i}}, {Iaria}, {Burderi},
  {Dov{\v{c}}iak}, {Karas}, {Matt}, {Papitto}, {Piraino}, {Riggio}, {Robba}, \&
  {Santangelo}}]{DiSalvo.etAl.2009}
{Di Salvo}, T., {D'A{\'\i}}, A., {Iaria}, R., {et~al.} 2009, \mnras, 398, 2022

\bibitem[{{Di Salvo} {et~al.}(2024){Di Salvo}, {Papitto}, {Marino}, {Iaria}, \&
  {Burderi}}]{DiSalvo.etAl.2023}
{Di Salvo}, T., {Papitto}, A., {Marino}, A., {Iaria}, R., \& {Burderi}, L.
  2024, in Handbook of X-ray and Gamma-ray Astrophysics, ed. C.~{Bambi} \&
  A.~{Santangelo} (Singapore: Springer), 4031--4103

\bibitem[{{Durant} {et~al.}(2010){Durant}, {Cornelisse}, {Remillard}, \&
  {Levine}}]{Durant.etAl.2010}
{Durant}, M., {Cornelisse}, R., {Remillard}, R., \& {Levine}, A. 2010, \mnras,
  401, 355

\bibitem[{{Farinelli} {et~al.}(2023){Farinelli}, {Fabiani}, {Poutanen},
  {Ursini}, {Ferrigno}, {Bianchi}, {Cocchi}, {Capitanio}, {De Rosa}, {Gnarini},
  {Kislat}, {Matt}, {Mikusincova}, {Muleri}, {Agudo}, {Antonelli}, {Bachetti},
  {Baldini}, {Baumgartner}, {Bellazzini}, {Bongiorno}, {Bonino}, {Brez},
  {Bucciantini}, {Castellano}, {Cavazzuti}, {Ciprini}, {Costa}, {Del Monte},
  {Di Gesu}, {Di Lalla}, {Di Marco}, {Donnarumma}, {Doroshenko},
  {Dov{\v{c}}iak}, {Ehlert}, {Enoto}, {Evangelista}, {Ferrazzoli}, {Garcia},
  {Gunji}, {Hayashida}, {Heyl}, {Iwakiri}, {Jorstad}, {Karas}, {Kitaguchi},
  {Kolodziejczak}, {Krawczynski}, {La Monaca}, {Latronico}, {Liodakis},
  {Maldera}, {Manfreda}, {Marin}, {Marscher}, {Marshall}, {Mitsuishi},
  {Mizuno}, {Ng}, {O'Dell}, {Omodei}, {Oppedisano}, {Papitto}, {Pavlov},
  {Peirson}, {Perri}, {Pesce-Rollins}, {Petrucci}, {Pilia}, {Possenti},
  {Puccetti}, {Ramsey}, {Rankin}, {Ratheesh}, {Romani}, {Sgr{\`o}}, {Slane},
  {Soffitta}, {Spandre}, {Tamagawa}, {Tavecchio}, {Taverna}, {Tawara},
  {Tennant}, {Thomas}, {Tombesi}, {Trois}, {Tsygankov}, {Turolla}, {Vink},
  {Weisskopf}, {Wu}, {Xie}, \& {Zane}}]{Farinelli.etAl.2023}
{Farinelli}, R., {Fabiani}, S., {Poutanen}, J., {et~al.} 2023, \mnras, 519,
  3681

\bibitem[{{Farinelli} {et~al.}(2024){Farinelli}, {Waghmare}, {Ducci}, \&
  {Santangelo}}]{Farinelli.etAl.2024}
{Farinelli}, R., {Waghmare}, A., {Ducci}, L., \& {Santangelo}, A. 2024, \aap,
  684, A62

\bibitem[{{Galloway} {et~al.}(2008){Galloway}, {Muno}, {Hartman}, {Psaltis}, \&
  {Chakrabarty}}]{Galloway.etAl.2008}
{Galloway}, D.~K., {Muno}, M.~P., {Hartman}, J.~M., {Psaltis}, D., \&
  {Chakrabarty}, D. 2008, \apjs, 179, 360

\bibitem[{{Garc{\'\i}a} {et~al.}(2014){Garc{\'\i}a}, {Dauser}, {Lohfink},
  {Kallman}, {Steiner}, {McClintock}, {Brenneman}, {Wilms}, {Eikmann},
  {Reynolds}, \& {Tombesi}}]{Garcia.etAl.2014}
{Garc{\'\i}a}, J., {Dauser}, T., {Lohfink}, A., {et~al.} 2014, \apj, 782, 76

\bibitem[{{Garc{\'\i}a} {et~al.}(2016){Garc{\'\i}a}, {Fabian}, {Kallman},
  {Dauser}, {Parker}, {McClintock}, {Steiner}, \& {Wilms}}]{Garcia.etAl.2016}
{Garc{\'\i}a}, J.~A., {Fabian}, A.~C., {Kallman}, T.~R., {et~al.} 2016, \mnras,
  462, 751

\bibitem[{{Gendreau} {et~al.}(2016){Gendreau}, {Arzoumanian}, {Adkins},
  {Albert}, {Anders}, {Aylward}, {Baker}, {Balsamo}, {Bamford}, {Benegalrao},
  {Berry}, {Bhalwani}, {Black}, {Blaurock}, {Bronke}, {Brown}, {Budinoff},
  {Cantwell}, {Cazeau}, {Chen}, {Clement}, {Colangelo}, {Coleman},
  {Coopersmith}, {Dehaven}, {Doty}, {Egan}, {Enoto}, {Fan}, {Ferro}, {Foster},
  {Galassi}, {Gallo}, {Green}, {Grosh}, {Ha}, {Hasouneh}, {Heefner}, {Hestnes},
  {Hoge}, {Jacobs}, {J{\o}rgensen}, {Kaiser}, {Kellogg}, {Kenyon}, {Koenecke},
  {Kozon}, {LaMarr}, {Lambertson}, {Larson}, {Lentine}, {Lewis}, {Lilly},
  {Liu}, {Malonis}, {Manthripragada}, {Markwardt}, {Matonak}, {Mcginnis},
  {Miller}, {Mitchell}, {Mitchell}, {Mohammed}, {Monroe}, {Montt de Garcia},
  {Mul{\'e}}, {Nagao}, {Ngo}, {Norris}, {Norwood}, {Novotka}, {Okajima},
  {Olsen}, {Onyeachu}, {Orosco}, {Peterson}, {Pevear}, {Pham}, {Pollard},
  {Pope}, {Powers}, {Powers}, {Price}, {Prigozhin}, {Ramirez}, {Reid},
  {Remillard}, {Rogstad}, {Rosecrans}, {Rowe}, {Sager}, {Sanders}, {Savadkin},
  {Saylor}, {Schaeffer}, {Schweiss}, {Semper}, {Serlemitsos}, {Shackelford},
  {Soong}, {Struebel}, {Vezie}, {Villasenor}, {Winternitz}, {Wofford},
  {Wright}, {Yang}, \& {Yu}}]{Gendreau.etAl.2016}
{Gendreau}, K.~C., {Arzoumanian}, Z., {Adkins}, P.~W., {et~al.} 2016, in
  \procspie, Vol. 9905, Space Telescopes and Instrumentation 2016: Ultraviolet
  to Gamma Ray, ed. J.-W.~A. {den Herder}, T.~{Takahashi}, \& M.~{Bautz},
  99051H

\bibitem[{Gnarini {et~al.}(2022)Gnarini, Ursini, Matt, Bianchi, Capitanio,
  Cocchi, Farinelli, \& Zhang}]{Gnarini.etAl.2022}
Gnarini, A., Ursini, F., Matt, G., {et~al.} 2022, \mnras, 514, 2561

\bibitem[{{Harrison} {et~al.}(2013){Harrison}, {Craig}, {Christensen},
  {Hailey}, {Zhang}, {Boggs}, {Stern}, {Cook}, {Forster}, {Giommi},
  {Grefenstette}, {Kim}, {Kitaguchi}, {Koglin}, {Madsen}, {Mao}, {Miyasaka},
  {Mori}, {Perri}, {Pivovaroff}, {Puccetti}, {Rana}, {Westergaard}, {Willis},
  {Zoglauer}, {An}, {Bachetti}, {Barri{\`e}re}, {Bellm}, {Bhalerao},
  {Brejnholt}, {Fuerst}, {Liebe}, {Markwardt}, {Nynka}, {Vogel}, {Walton},
  {Wik}, {Alexander}, {Cominsky}, {Hornschemeier}, {Hornstrup}, {Kaspi},
  {Madejski}, {Matt}, {Molendi}, {Smith}, {Tomsick}, {Ajello}, {Ballantyne},
  {Balokovi{\'c}}, {Barret}, {Bauer}, {Blandford}, {Brandt}, {Brenneman},
  {Chiang}, {Chakrabarty}, {Chenevez}, {Comastri}, {Dufour}, {Elvis}, {Fabian},
  {Farrah}, {Fryer}, {Gotthelf}, {Grindlay}, {Helfand}, {Krivonos}, {Meier},
  {Miller}, {Natalucci}, {Ogle}, {Ofek}, {Ptak}, {Reynolds}, {Rigby},
  {Tagliaferri}, {Thorsett}, {Treister}, \& {Urry}}]{Harrison.etAl.2013}
{Harrison}, F.~A., {Craig}, W.~W., {Christensen}, F.~E., {et~al.} 2013, \apj,
  770, 103

\bibitem[{{Hasinger} \& {van der Klis}(1989)}]{Hasinger.VanDerKlis.1989}
{Hasinger}, G. \& {van der Klis}, M. 1989, \aap, 225, 79

\bibitem[{{Iaria} {et~al.}(2020){Iaria}, {Mazzola}, {Di Salvo}, {Marino},
  {Gambino}, {Sanna}, {Riggio}, \& {Burderi}}]{Iaria.etAl.2020}
{Iaria}, R., {Mazzola}, S.~M., {Di Salvo}, T., {et~al.} 2020, \aap, 635, A209

\bibitem[{{Inogamov} \& {Sunyaev}(1999)}]{Inogamov.Sunyaev.1999}
{Inogamov}, N.~A. \& {Sunyaev}, R.~A. 1999, Astronomy Letters, 25, 269

\bibitem[{{Kaastra} \& {Bleeker}(2016)}]{Kaastra.Bleeker.2016}
{Kaastra}, J.~S. \& {Bleeker}, J. A.~M. 2016, \aap, 587, A151

\bibitem[{{Kotze} \& {Charles}(2010)}]{Kotze.Charles.2010}
{Kotze}, M.~M. \& {Charles}, P.~A. 2010, \mnras, 402, L16

\bibitem[{{Kuulkers}(2002)}]{Kuulkers.2002}
{Kuulkers}, E. 2002, \aap, 383, L5

\bibitem[{{Kuulkers} \& {van der Klis}(2000)}]{Kuulkers.VanDerKlis.2000}
{Kuulkers}, E. \& {van der Klis}, M. 2000, \aap, 356, L45

\bibitem[{{Kuulkers} {et~al.}(1994){Kuulkers}, {van der Klis}, {Oosterbroek},
  {Asai}, {Dotani}, {van Paradijs}, \& {Lewin}}]{Kuulkers.etAl.1994}
{Kuulkers}, E., {van der Klis}, M., {Oosterbroek}, T., {et~al.} 1994, \aap,
  289, 795

\bibitem[{{La Monaca} {et~al.}(2024){La Monaca}, {Di Marco}, {Poutanen},
  {Bachetti}, {Motta}, {Papitto}, {Pilia}, {Xie}, {Bianchi}, {Bobrikova},
  {Costa}, {Deng}, {Ge}, {Illiano}, {Jia}, {Krawczynski}, {Lai}, {Liu},
  {Mastroserio}, {Muleri}, {Rankin}, {Soffitta}, {Veledina}, {Ambrosino}, {Del
  Santo}, {Chen}, {Garcia}, {Kaaret}, {Russell}, {Wei}, {Zhang}, {Zuo},
  {Arzoumanian}, {Cocchi}, {Gnarini}, {Farinelli}, {Gendreau}, {Ursini},
  {Weisskopf}, {Zane}, {Agudo}, {Antonelli}, {Baldini}, {Baumgartner},
  {Bellazzini}, {Bongiorno}, {Bonino}, {Brez}, {Bucciantini}, {Capitanio},
  {Castellano}, {Cavazzuti}, {Chen}, {Ciprini}, {De Rosa}, {Del Monte}, {Di
  Gesu}, {Di Lalla}, {Donnarumma}, {Doroshenko}, {Dovciak}, {Ehlert}, {Enoto},
  {Evangelista}, {Fabiani}, {Ferrazzoli}, {Gunji}, {Hayashidadag}, {Heyl},
  {Iwakiri}, {Jorstad}, {Karas}, {Kislat}, {Kitaguchi}, {Kolodziejczak},
  {Latronico}, {Liodakis}, {Maldera}, {Manfreda}, {Marin}, {Marinucci},
  {Marscher}, {Marshall}, {Massaro}, {Matt}, {Mitsuishi}, {Mizuno}, {Negro},
  {Chi-Yung Ng}, {O'Dell}, {Omodei}, {Oppedisano}, {Pavlov}, {Peirson},
  {Perri}, {Pesce-Rollins}, {Petrucci}, {Possenti}, {Puccetti}, {Ramsey},
  {Ratheesh}, {Roberts}, {Romani}, {Sgr{\`o}}, {Slane}, {Spandre}, {Swartz},
  {Tamagawa}, {Tavecchio}, {Taverna}, {Tawara}, {Tennant}, {Thomas}, {Tombesi},
  {Trois}, {Tsygankov}, {Turolla}, {Vink}, \& {Wu}}]{LaMonaca.etAl.2024}
{La Monaca}, F., {Di Marco}, A., {Poutanen}, J., {et~al.} 2024, \apjl, 960, L11

\bibitem[{{Lapidus} \& {Sunyaev}(1985)}]{Lapidus.Sunyaev.1985}
{Lapidus}, I.~I. \& {Sunyaev}, R.~A. 1985, \mnras, 217, 291

\bibitem[{{Lewin} {et~al.}(1987){Lewin}, {van Paradijs}, {Hasinger}, {Penninx},
  {Langmeier}, {van der Klis}, {Jansen}, {Basinska}, {Sztajno}, \&
  {Trumper}}]{Lewin.etAl.1987}
{Lewin}, W.~H.~G., {van Paradijs}, J., {Hasinger}, G., {et~al.} 1987, \mnras,
  226, 383

\bibitem[{{Ludlam}(2024)}]{Ludlam.2024}
{Ludlam}, R.~M. 2024, \apss, 369, 16

\bibitem[{Ludlam {et~al.}(2022)Ludlam, Cackett, Garc{\'{\i}}a, Miller, Stevens,
  Fabian, Homan, Ng, Guillot, Buisson, \& Chakrabarty}]{Ludlam.etAl.2022}
Ludlam, R.~M., Cackett, E.~M., Garc{\'{\i}}a, J.~A., {et~al.} 2022, \apj, 927,
  112

\bibitem[{{Ludlam} {et~al.}(2019){Ludlam}, {Miller}, {Barret}, {Cackett},
  {Coughenour}, {Dauser}, {Degenaar}, {Garc{\'\i}a}, {Harrison}, \&
  {Paerels}}]{Ludlam.etAl.2019}
{Ludlam}, R.~M., {Miller}, J.~M., {Barret}, D., {et~al.} 2019, \apj, 873, 99

\bibitem[{{Mainardi} {et~al.}(2010){Mainardi}, {Paizis}, {Farinelli},
  {Kuulkers}, {Rodriguez}, {Hannikainen}, {Savolainen}, {Piraino}, {Bazzano},
  \& {Santangelo}}]{Mainardi.etAl.2010}
{Mainardi}, L.~I., {Paizis}, A., {Farinelli}, R., {et~al.} 2010, \aap, 512, A57

\bibitem[{{Makishima} {et~al.}(1983){Makishima}, {Mitsuda}, {Inoue}, {Koyama},
  {Matsuoka}, {Murakami}, {Oda}, {Ogawara}, {Ohashi}, {Shibazaki}, {Tanaka},
  {Marshall}, {Hayakawa}, {Kunieda}, {Makino}, {Nagase}, {Tawara}, {Miyamoto},
  {Tsunemi}, {Tsuno}, {Yamashita}, \& {Kondo}}]{Makishima.etAl.1983}
{Makishima}, K., {Mitsuda}, K., {Inoue}, H., {et~al.} 1983, \apj, 267, 310

\bibitem[{{Marinucci} {et~al.}(2022){Marinucci}, {Muleri}, {Dovciak},
  {Bianchi}, {Marin}, {Matt}, {Ursini}, {Middei}, {Marshall}, {Baldini},
  {Barnouin}, {Rodriguez}, {De Rosa}, {Di Gesu}, {Harper}, {Ingram}, {Karas},
  {Krawczynski}, {Madejski}, {Panagiotou}, {Petrucci}, {Podgorny}, {Puccetti},
  {Tombesi}, {Veledina}, {Zhang}, {Agudo}, {Antonelli}, {Bachetti},
  {Baumgartner}, {Bellazzini}, {Bongiorno}, {Bonino}, {Brez}, {Bucciantini},
  {Capitanio}, {Castellano}, {Cavazzuti}, {Ciprini}, {Costa}, {Del Monte}, {Di
  Lalla}, {Di Marco}, {Donnarumma}, {Doroshenko}, {Ehlert}, {Enoto},
  {Evangelista}, {Fabiani}, {Ferrazzoli}, {Garcia}, {Gunji}, {Hayashida},
  {Heyl}, {Iwakiri}, {Jorstad}, {Kitaguchi}, {Kolodziejczak}, {La Monaca},
  {Latronico}, {Liodakis}, {Maldera}, {Manfreda}, {Marscher}, {Mitsuishi},
  {Mizuno}, {Ng}, {O'Dell}, {Omodei}, {Oppedisano}, {Papitto}, {Pavlov},
  {Peirson}, {Perri}, {Pesce-Rollins}, {Pilia}, {Possenti}, {Poutanen},
  {Ramsey}, {Rankin}, {Ratheesh}, {Romani}, {Sgr{\v{s}}}, {Slane}, {Soffitta},
  {Spandre}, {Tamagawa}, {Tavecchio}, {Taverna}, {Tawara}, {Tennant}, {Thomas},
  {Trois}, {Tsygankov}, {Turolla}, {Vink}, {Weisskopf}, {Wu}, {Xie}, \&
  {Zane}}]{Marinucci.etAl.2022}
{Marinucci}, A., {Muleri}, F., {Dovciak}, M., {et~al.} 2022, \mnras, 516, 5907

\bibitem[{{Matt}(1993)}]{Matt.1993}
{Matt}, G. 1993, \mnras, 260, 663

\bibitem[{{Migliari} \& {Fender}(2006)}]{Migliari.Fender.2006}
{Migliari}, S. \& {Fender}, R.~P. 2006, \mnras, 366, 79

\bibitem[{{Miller} {et~al.}(2018){Miller}, {Gendreau}, {Ludlam}, {Fabian},
  {Altamirano}, {Arzoumanian}, {Bult}, {Cackett}, {Homan}, {Kara}, {Neilsen},
  {Remillard}, \& {Tombesi}}]{Miller.etAl.2018}
{Miller}, J.~M., {Gendreau}, K., {Ludlam}, R.~M., {et~al.} 2018, \apjl, 860,
  L28

\bibitem[{{Miller} {et~al.}(2013){Miller}, {Parker}, {Fuerst}, {Bachetti},
  {Barret}, {Grefenstette}, {Tendulkar}, {Harrison}, {Boggs}, {Chakrabarty},
  {Christensen}, {Craig}, {Fabian}, {Hailey}, {Natalucci}, {Paerels}, {Rana},
  {Stern}, {Tomsick}, \& {Zhang}}]{Miller.etAl.2013}
{Miller}, J.~M., {Parker}, M.~L., {Fuerst}, F., {et~al.} 2013, \apjl, 779, L2

\bibitem[{{Mitsuda} {et~al.}(1984){Mitsuda}, {Inoue}, {Koyama}, {Makishima},
  {Matsuoka}, {Ogawara}, {Shibazaki}, {Suzuki}, {Tanaka}, \&
  {Hirano}}]{Mitsuda.etAl.1984}
{Mitsuda}, K., {Inoue}, H., {Koyama}, K., {et~al.} 1984, \pasj, 36, 741

\bibitem[{{Mitsuda} {et~al.}(1989){Mitsuda}, {Inoue}, {Nakamura}, \&
  {Tanaka}}]{Mitsuda.etAl.1989}
{Mitsuda}, K., {Inoue}, H., {Nakamura}, N., \& {Tanaka}, Y. 1989, \pasj, 41, 97

\bibitem[{{Mondal} {et~al.}(2017){Mondal}, {Pahari}, {Dewangan}, {Misra}, \&
  {Raychaudhuri}}]{Mondal.etAl.2017}
{Mondal}, A.~S., {Pahari}, M., {Dewangan}, G.~C., {Misra}, R., \&
  {Raychaudhuri}, B. 2017, \mnras, 466, 4991

\bibitem[{{Nasa High Energy Astrophysics Science Archive Research
  Center}(2014)}]{Heasarc}
{Nasa High Energy Astrophysics Science Archive Research Center}. 2014,
  {HEAsoft: Unified Release of FTOOLS and XANADU}, Astrophysics Source Code
  Library, record ascl:1408.004

\bibitem[{{Oosterbroek} {et~al.}(2001){Oosterbroek}, {Barret}, {Guainazzi}, \&
  {Ford}}]{Oosterbroek.etAl.2001}
{Oosterbroek}, T., {Barret}, D., {Guainazzi}, M., \& {Ford}, E.~C. 2001, \aap,
  366, 138

\bibitem[{Patruno {et~al.}(2017)Patruno, Haskell, \& Andersson}]{Patruno.2017}
Patruno, A., Haskell, B., \& Andersson, N. 2017, \apj, 850, 106

\bibitem[{{Piconcelli} {et~al.}(2004){Piconcelli}, {Jimenez-Bail{\'o}n},
  {Guainazzi}, {Schartel}, {Rodr{\'\i}guez-Pascual}, \&
  {Santos-Lle{\'o}}}]{Piconcelli.etAl.2004}
{Piconcelli}, E., {Jimenez-Bail{\'o}n}, E., {Guainazzi}, M., {et~al.} 2004,
  \mnras, 351, 161

\bibitem[{{Pintore} {et~al.}(2015){Pintore}, {Di Salvo}, {Bozzo}, {Sanna},
  {Burderi}, {D'A{\`\i}}, {Riggio}, {Scarano}, \& {Iaria}}]{Pintore.etAl.2015}
{Pintore}, F., {Di Salvo}, T., {Bozzo}, E., {et~al.} 2015, \mnras, 450, 2016

\bibitem[{{Piraino} {et~al.}(2012){Piraino}, {Santangelo}, {Kaaret},
  {M{\"u}ck}, {D'A{\`\i}}, {Di Salvo}, {Iaria}, {Robba}, {Burderi}, \&
  {Egron}}]{Piraino.etAl.2012}
{Piraino}, S., {Santangelo}, A., {Kaaret}, P., {et~al.} 2012, \aap, 542, L27

\bibitem[{{Popham} \& {Sunyaev}(2001)}]{Popham.Sunyaev.2001}
{Popham}, R. \& {Sunyaev}, R. 2001, \apj, 547, 355

\bibitem[{{Poutanen} {et~al.}(1996){Poutanen}, {Nagendra}, \&
  {Svensson}}]{Poutanen.etAl.1996}
{Poutanen}, J., {Nagendra}, K.~N., \& {Svensson}, R. 1996, \mnras, 283, 892

\bibitem[{{Saade} {et~al.}(2024){Saade}, {Kaaret}, {Gnarini}, {Poutanen},
  {Ursini}, {Bianchi}, {Bobrikova}, {La Monaca}, {Di Marco}, {Capitanio},
  {Veledina}, {Agudo}, {Antonelli}, {Bachetti}, {Baldini}, {Baumgartner},
  {Bellazzini}, {Bongiorno}, {Bonino}, {Brez}, {Bucciantini}, {Castellano},
  {Cavazzuti}, {Chen}, {Ciprini}, {Costa}, {De Rosa}, {Del Monte}, {Di
  Ges{\`u}}, {Di Lalla}, {Donnarumma}, {Doroshenko}, {Dovciak}, {Ehlert},
  {Emote}, {Evangelista}, {Fabiani}, {Ferrazzoli}, {Garcia}, {Gunji},
  {Hayashida}, {Heyl}, {Iwakiri}, {Jorstad}, {Karas}, {Kislat}, {Kitaguchi},
  {Kolodziejczak}, {Krawczynski}, {Latronico}, {Liodakis}, {Maldera},
  {Manfreda}, {Marin}, {Marinucci}, {Marscher}, {Marshall}, {Massaro}, {Matt},
  {Mitsuishi}, {Mizudo}, {Muleri}, {Negro}, {Ng}, {O'Dell}, {Omodei},
  {Oppedisano}, {Papitto}, {Pavlov}, {Peirson}, {Perri}, {Pesce-Rollins},
  {Petrucci}, {Pilia}, {Possenti}, {Puccetti}, {Ramsey}, {Rankin}, {Ratheesh},
  {Roberts}, {Romani}, {Sgro}, {Slane}, {Soffitta}, {Spandre}, {Zwartz},
  {Tamagawa}, {Tavecchio}, {Taverna}, {Tawara}, {Tenant}, {Thomas}, {Tombesi},
  {Trois}, {Tsygankov}, {Turolla}, {Vink}, {Weisskopf}, {Wu}, {Xie}, \&
  {Zane}}]{Saade.etAl.2024}
{Saade}, M.~L., {Kaaret}, P., {Gnarini}, A., {et~al.} 2024, \apj, 963, 133

\bibitem[{{Seifina} \& {Titarchuk}(2012)}]{Seifina.Titarchuk.2012}
{Seifina}, E. \& {Titarchuk}, L. 2012, \apj, 747, 99

\bibitem[{{Strohmayer} {et~al.}(2018){Strohmayer}, {Gendreau}, {Altamirano},
  {Arzoumanian}, {Bult}, {Chakrabarty}, {Chenevez}, {Guillot}, {Guver},
  {Homan}, {Jaisawal}, {Keek}, {Mahmoodifar}, {Miller}, \&
  {Ozel}}]{Strohmayer.etAl.2018}
{Strohmayer}, T.~E., {Gendreau}, K.~C., {Altamirano}, D., {et~al.} 2018, \apj,
  865, 63

\bibitem[{Suleimanov \& Poutanen(2006)}]{Suleimanov.Poutanen.2006}
Suleimanov, V. \& Poutanen, J. 2006, \mnras, 369, 2036

\bibitem[{{Tarana} {et~al.}(2011){Tarana}, {Belloni}, {Bazzano}, {M{\'e}ndez},
  \& {Ubertini}}]{Tarana.etAl.2011}
{Tarana}, A., {Belloni}, T., {Bazzano}, A., {M{\'e}ndez}, M., \& {Ubertini}, P.
  2011, \mnras, 416, 873

\bibitem[{{Thomas} {et~al.}(2023){Thomas}, {Gudennavar}, \&
  {Bubbly}}]{Thomas.etAl.2023}
{Thomas}, N.~T., {Gudennavar}, S.~B., \& {Bubbly}, S.~G. 2023, \mnras, 521, 433

\bibitem[{{Ursini} {et~al.}(2023{\natexlab{a}}){Ursini}, {Farinelli},
  {Gnarini}, {Poutanen}, {Bianchi}, {Capitanio}, {Di Marco}, {Fabiani}, {La
  Monaca}, {Malacaria}, {Matt}, {Miku{\v{s}}incov{\'a}}, {Cocchi}, {Kaaret},
  {Kajava}, {Pilia}, {Zhang}, {Agudo}, {Antonelli}, {Bachetti}, {Baldini},
  {Baumgartner}, {Bellazzini}, {Bongiorno}, {Bonino}, {Brez}, {Bucciantini},
  {Castellano}, {Cavazzuti}, {Chen}, {Ciprini}, {Costa}, {De Rosa}, {Del
  Monte}, {Di Gesu}, {Di Lalla}, {Donnarumma}, {Doroshenko}, {Dov{\v{c}}iak},
  {Ehlert}, {Enoto}, {Evangelista}, {Ferrazzoli}, {Garcia}, {Gunji},
  {Hayashida}, {Heyl}, {Iwakiri}, {Jorstad}, {Karas}, {Kislat}, {Kitaguchi},
  {Kolodziejczak}, {Krawczynski}, {Latronico}, {Liodakis}, {Maldera},
  {Manfreda}, {Marin}, {Marinucci}, {Marscher}, {Marshall}, {Massaro},
  {Mitsuishi}, {Mizuno}, {Muleri}, {Negro}, {Ng}, {O'Dell}, {Omodei},
  {Oppedisano}, {Papitto}, {Pavlov}, {Peirson}, {Perri}, {Pesce-Rollins},
  {Petrucci}, {Pilia}, {Possenti}, {Puccetti}, {Ramsey}, {Rankin}, {Ratheesh},
  {Roberts}, {Romani}, {Sgr{\`o}}, {Slane}, {Soffitta}, {Spandre}, {Swartz},
  {Tamagawa}, {Tavecchio}, {Taverna}, {Tawara}, {Tennant}, {Thomas}, {Tombesi},
  {Trois}, {Tsygankov}, {Turolla}, {Vink}, {Weisskopf}, {Wu}, {Xie}, \&
  {Zane}}]{Ursini.etAl.2023}
{Ursini}, F., {Farinelli}, R., {Gnarini}, A., {et~al.} 2023{\natexlab{a}},
  \aap, 676, A20

\bibitem[{{Ursini} {et~al.}(2024{\natexlab{a}}){Ursini}, {Gnarini}, {Bianchi},
  {Bobrikova}, {Capitanio}, {Cocchi}, {Fabiani}, {Farinelli}, {Kaaret}, {Matt},
  {Ng}, {Poutanen}, {Ravi}, \& {Tarana}}]{Ursini.etAl.2024}
{Ursini}, F., {Gnarini}, A., {Bianchi}, S., {et~al.} 2024{\natexlab{a}}, \aap,
  690, A200

\bibitem[{{Ursini} {et~al.}(2024{\natexlab{b}}){Ursini}, {Gnarini},
  {Capitanio}, {Bobrikova}, {Cocchi}, {Di Marco}, {Fabiani}, {Farinelli}, {La
  Monaca}, {Rankin}, {Saade}, \& {Poutanen}}]{Ursini.2024.Review}
{Ursini}, F., {Gnarini}, A., {Capitanio}, F., {et~al.} 2024{\natexlab{b}},
  Galaxies, 12

\bibitem[{{Ursini} {et~al.}(2023{\natexlab{b}}){Ursini}, {Marinucci}, {Matt},
  {Bianchi}, {Marin}, {Marshall}, {Middei}, {Poutanen}, {Rogantini}, {De Rosa},
  {Di Gesu}, {Garc{\'\i}a}, {Ingram}, {Kim}, {Krawczynski}, {Puccetti},
  {Soffitta}, {Svoboda}, {Tombesi}, {Weisskopf}, {Barnouin}, {Perri},
  {Podgorny}, {Ratheesh}, {Zaino}, {Agudo}, {Antonelli}, {Bachetti}, {Baldini},
  {Baumgartner}, {Bellazzini}, {Bongiorno}, {Bonino}, {Brez}, {Bucciantini},
  {Capitanio}, {Castellano}, {Cavazzuti}, {Ciprini}, {Costa}, {Del Monte}, {Di
  Lalla}, {Di Marco}, {Donnarumma}, {Doroshenko}, {Dovciak}, {Ehlert}, {Enoto},
  {Evangelista}, {Fabiani}, {Ferrazzoli}, {Gunji}, {Heyl}, {Iwakiri},
  {Jorstad}, {Karas}, {Kitaguchi}, {Kolodziejczak}, {La Monaca}, {Latronico},
  {Liodakis}, {Maldera}, {Manfreda}, {Marscher}, {Mitsuishi}, {Mizuno},
  {Muleri}, {Ng}, {O'Dell}, {Omodei}, {Oppedisano}, {Papitto}, {Pavlov},
  {Peirson}, {Pesce-Rollins}, {Petrucci}, {Pilia}, {Possenti}, {Ramsey},
  {Rankin}, {Romani}, {Sgr{\`o}}, {Slane}, {Spandre}, {Tamagawa}, {Tavecchio},
  {Taverna}, {Tawara}, {Tennant}, {Thomas}, {Trois}, {Tsygankov}, {Turolla},
  {Vink}, {Wu}, {Xie}, \& {Zane}}]{Ursini.etAl.2023.CirGal}
{Ursini}, F., {Marinucci}, A., {Matt}, G., {et~al.} 2023{\natexlab{b}}, \mnras,
  519, 50

\bibitem[{{van den Berg} {et~al.}(2014){van den Berg}, {Homan}, {Fridriksson},
  \& {Linares}}]{VanDenBerg.etAl.2014}
{van den Berg}, M., {Homan}, J., {Fridriksson}, J.~K., \& {Linares}, M. 2014,
  \apj, 793, 128

\bibitem[{{van der Klis}(1989)}]{VanDerKlis.1989}
{van der Klis}, M. 1989, \araa, 27, 517

\bibitem[{van~der Klis(1995)}]{VanDerKlis.1995}
van~der Klis, M. 1995, in Cambridge Astrophysics Series, Vol.~26, X-ray
  Binaries, ed. W.~H. Lewin, E.~P. van~den Heuvel, \& J.~van Paradijs
  (Cambridge: Cambridge University Press), 252--307

\bibitem[{van~der Klis(2006)}]{VanDerKlis.2006}
van~der Klis, M. 2006, in Cambridge Astrophysics Series, Vol.~39, Compact
  stellar X-ray sources, ed. W.~Lewin \& M.~van~der Klis (Cambridge: Cambridge
  University Press), 39–112

\bibitem[{{Verner} {et~al.}(1996){Verner}, {Ferland}, {Korista}, \&
  {Yakovlev}}]{Verner.etAl.1996}
{Verner}, D.~A., {Ferland}, G.~J., {Korista}, K.~T., \& {Yakovlev}, D.~G. 1996,
  \apj, 465, 487

\bibitem[{{Weisskopf} {et~al.}(2016){Weisskopf}, {Ramsey}, {O'Dell}, {Tennant},
  {Elsner}, {Soffitta}, {Bellazzini}, {Costa}, {Kolodziejczak}, {Kaspi},
  {Muleri}, {Marshall}, {Matt}, \& {Romani}}]{Weisskopf.etAl.2016}
{Weisskopf}, M.~C., {Ramsey}, B., {O'Dell}, S., {et~al.} 2016, in \procspie,
  Vol. 9905, Space Telescopes and Instrumentation 2016: Ultraviolet to Gamma
  Ray, ed. J.-W.~A. {den Herder}, T.~{Takahashi}, \& M.~{Bautz}, 990517

\bibitem[{Weisskopf {et~al.}(2022)Weisskopf, Soffitta, Baldini, Ramsey, O'Dell,
  Romani, Matt, Deininger, Baumgartner, Bellazzini, Costa, Kolodziejczak,
  Latronico, Marshall, Muleri, Bongiorno, Tennant, Bucciantini, Dovciak, Marin,
  Marscher, Poutanen, Slane, Turolla, Kalinowski, Marco, Fabiani, Minuti,
  Monaca, Pinchera, Rankin, Sgrò, Trois, Xie, Alexander, Allen, Amici,
  Andersen, Antonelli, Antoniak, Attiná, Barbanera, Bachetti, Baggett, Bladt,
  Brez, Bonino, Boree, Borotto, Breeding, Brienza, Bygott, Caporale, Cardelli,
  Carpentiero, Castellano, Castronuovo, Cavalli, Cavazzuti, Ceccanti, Centrone,
  Citraro, D'Amico, D'Alba, Gesu, Monte, Dietz, Lalla, Persio, Dolan,
  Donnarumma, Evangelista, Ferrant, Ferrazzoli, Ferrie, Footdale, Forsyth,
  Foster, Garelick, Gunji, Gurnee, Head, Hibbard, Johnson, Kelly, Kilaru,
  Lefevre, Roy, Loffredo, Lorenzi, Lucchesi, Maddox, Magazzu, Maldera,
  Manfreda, Mangraviti, Marengo, Marrocchesi, Massaro, Mauger, McCracken,
  McEachen, Mize, Mereu, Mitchell, Mitsuishi, Morbidini, Mosti, Nasimi, Negri,
  Negro, Nguyen, Nitschke, Nuti, Onizuka, Oppedisano, Orsini, Osborne, Pacheco,
  Paggi, Painter, Pavelitz, Pentz, Piazzolla, Perri, Pesce-Rollins, Peterson,
  Pilia, Profeti, Puccetti, Ranganathan, Ratheesh, Reedy, Root, Rubini,
  Ruswick, Sanchez, Sarra, Santoli, Scalise, Sciortino, Schroeder, Seek,
  Sosdian, Spandre, Speegle, Tamagawa, Tardiola, Tobia, Thomas, Valerie,
  Vimercati, Walden, Weddendorf, Wedmore, Welch, Zanetti, \&
  Zanetti}]{Weisskopf.2022}
Weisskopf, M.~C., Soffitta, P., Baldini, L., {et~al.} 2022, JATIS, 8, 1

\bibitem[{{Wilms} {et~al.}(2000){Wilms}, {Allen}, \&
  {McCray}}]{Wilms.etAl.2000}
{Wilms}, J., {Allen}, A., \& {McCray}, R. 2000, \apj, 542, 914

\bibitem[{{Zdziarski} {et~al.}(2020){Zdziarski}, {Szanecki}, {Poutanen},
  {Gierli{\'n}ski}, \& {Biernacki}}]{Zdziarski.etAl.2020}
{Zdziarski}, A.~A., {Szanecki}, M., {Poutanen}, J., {Gierli{\'n}ski}, M., \&
  {Biernacki}, P. 2020, \mnras, 492, 5234

\end{thebibliography}

\end{document}